\documentclass[a4paper,UKenglish]{lipicsown}

\def\arxivy{yes}
\def\noy{no}

\usepackage{url}
\usepackage{color}
\usepackage[usenames,dvipsnames,svgnames,table]{xcolor}
\usepackage{hyperref}
\hypersetup{colorlinks   = true,
     citecolor    = blue,
     urlcolor     = violet,
     linkcolor    = blue}
\def\ETH{Exponential Time Hypothesis}
\usepackage{tikz}
\usetikzlibrary{arrows,automata}

\usepackage{amsfonts}
\usepackage{amssymb}
\usepackage{amsmath}
\usepackage{mathtools}
\usepackage{float}
\usepackage{tikz}

\newtheorem{propo}[theorem]{Proposition}
\newtheorem{claim}[theorem]{Claim}

\newlength\myindent 
\newcommand\polylog{{\rm polylog}} 
\newcommand\Lang{{\rm Lang}} 
\usepackage{enumerate}

\DeclarePairedDelimiter\ceil{\lceil}{\rceil}

\title{Languages given by Finite Automata over the Unary Alphabet}


\author[5]{Wojciech Czerwi\'nski$^1$, Maciej D\k ebski$^2$, Tomasz Gogasz$^1$,
  \newline Gordon Hoi$^3$, Sanjay Jain$^4$, Micha\l\ Skrzypczak$^1$, \newline
  Frank Stephan$^{4,5}$ and Christopher Tan}
\affil[1]{Institute of Informatics, Faculty of Mathematics,
  Informatics and Mechanics, University of Warsaw,
  ulica Banacha 2, 02-097 Warsaw, Poland,
\texttt{wczerwin@mimuw.edu.pl, t.gogacz@mimuw.edu.pl,
  mskrzypczak@mimuw.edu.pl};
W. Czerwi\'nski was supported by the
ERC grant INFSYS, agreement no.\ 950398.
M. Skyrzypczak was supported by the
National Science Centre, Poland (grant no.
2021/41/B/ST6/03914).}
\affil[2]{Warsaw, Poland, \texttt{mdempsej@gmail.com}.}
\affil[3]{School of Informatics and IT, Temasek Polytechnic,
21 Tampines Ave 1, Singapore 529757, Republic of Singapore,
\texttt{hoickg@gmail.com}.}
\affil[4]{School of Computing, National University of Singapore,
13 Computing Drive, Block COM1, Singapore 117417,
Republic of Singapore, \texttt{sanjay@comp.nus.edu.sg}; 
S.~Jain was supported in part by the Singapore Ministry of Education
Academic Research Fund Tier 2 grants MOE2019-T2-2-121 / R146-000-304-112
and MOE-000538-00
as well as NUS Provost Chair grant C252-000-087-001 / E-252-00-0021-01.}
\affil[5]{Department of Mathematics,
National University of Singapore,
10 Lower Kent Ridge Road, Block S17, Singapore 119076,
Republic of Singapore, \texttt{fstephan@comp.nus.edu.sg} and
\texttt{e0774066@u.nus.edu};
F.~Stephan was supported in part by the Singapore Ministry of Education
Academic Research Fund Tier 2 grants MOE2019-T2-2-121 / R146-000-304-112
and MOE-000538-00.}

\authorrunning{Czerwi\'nski, D\k ebski, Gogasz,
  Hoi, Jain, Skrzypczak, Stephan and Tan}
\titlerunning{Languages and Automata for the Unary Alphabet}
\date{\today}

\begin{document}
\maketitle

\begin{abstract}
This paper studies the complexity of operations on finite automata and
the complexity of their decision problems when the alphabet is unary.
Let $n$ denote the maximum of the number of states of the 
input finite automata considered in the corresponding results.
The following main results are obtained:

(1) Given two unary NFAs recognising $L$ and $H$, respectively,
one can decide whether $L \subseteq H$ as well as
whether $L = H$ in time $2^{O((n \log n)^{1/3})}$. 
The previous upper bound on time was $2^{O((n \log n)^{1/2})}$
as given by Chrobak (1986),
and this bound was not significantly improved since then.

(2) Given two unary UFAs (unambiguous finite automata) recognising
$L$ and $H$, respectively, one can determine a UFA recognising
$L \cup H$ and a UFA recognising complement of $L$,
where these output UFAs have the number
of states bounded by a quasipolynomial in $n$.
However, in the worst case, a UFA for recognising concatenation of
languages recognised by two $n$-state UFAs,
uses $2^{\Theta((n \log^2 n)^{1/3})}$ states.

(3) Given a unary language $L$, if $L$ contains the word of length $k$,
then let $L(k)=1$ else let $L(k)=0$.
Let $\omega_L$ be the $\omega$-word
$L(0)L(1)\ldots$ and let $\cal L$ be a fixed $\omega$-regular language.
The last section studies how difficult it is to decide,
given an $n$-state UFA or NFA
recognising some language $L$, whether $\omega_L \in {\cal L}$.
\end{abstract}

\section{Introduction}  \label{sec:intro}

\noindent
This paper investigates the complexity- and size-constraints related
to languages over the unary alphabet -- this is assumed everywhere
throughout the paper -- when these languages are given by a nondeterministic
finite automaton (NFA) with a special emphasis on the case where this NFA
is unambiguous. Unambiguous nondeterministic finite automata (UFAs)
have many good algorithmic properties, even under regular operations
with the languages, as long as no concatenation is involved.
The study of unary languages is a theoretically important special case which
allows for techniques and insights from number theory, as each word
corresponds to the natural number which is its length. This sometimes
gives techniques which can be extended to the general case where alphabet size
is larger.
Furthermore, the case is a bit special as the NFA-DFA trade-off
is only $2^{O((n \log n)^{1/2})}$ and not $2^n$ as for
binary alphabets. Colcombet \cite{Co15} published several influential
conjectures which found much attention. G\"o\"os, Kiefer and Yuan
\cite{GKY22} refuted Colcombet's conjecture that
any two $n$-state NFAs with disjoint languages can be separated
by a UFA of size polynomially in $n$. 
Raskin \cite{Ra18} had refuted another conjecture
of Colcombet about polynomial blow-up for Boolean operations for
UFAs.
In this paper, a weaker version of this conjecture by Colcombet is proven:
unary UFAs have only a quasipolynomial blow-up for Boolean operations.

In the following, a bound of type $2^{\Theta(n)}$ is called exponential
and a bound of type $2^{n^{\Theta(1)}}$ is called exponential-type.
The quasipolynomial functions are those in $O(n^{\log^{O(1)}(n)})$ which
are not bounded by polynomials.
In an expression of the form $2^{\alpha(n)}$, the function
$\alpha(n)$ is called the exponent of the function.

In the following, let $f \in \Omega'(g)$ mean that there is a constant
$c>0$ such that, for infinitely many $n$, $f(n) \geq c \cdot g(n)$.
Note that, when such a lower bound involving $\Omega'(g)$
is used in this paper, it means that, for some fixed constant $c$,
for any algorithm solving the corresponding problems, the lower
bound applies for infinitely many $n$.

Note that, under the Exponential Time Hypothesis (ETH),
all \textbf{NP}-complete
problems have, for infinitely many inputs, an exponential-type complexity
and solving $k$SAT with $k \geq 3$ requires time $2^{\Omega'(n)}$.
In this paper, for several results, assuming \ETH, 
the exponents of the exponential
type function in the lower and upper bounds are matched within
a logarithmic or sublogarithmic function.

For the ease of presentation, when dealing with some operations
(such as union, intersection, complementation etc) on
languages recognised by NFAs,
it may be written as the operations on the NFAs.

Universality problem for an automaton is to check if
the language recognised by it is equal to $\Sigma^*$.
Comparison problem is to check, given two automaton $N_1$ and $N_2$,
if the language recognised by $N_1$
is contained in the language recognised by $N_2$.
Assume that the NFAs given are of at most $n$-states.
Fernau and Krebs \cite{FK17} proved, under the assumption of
ETH, the conditional lower bound of
$2^{\Omega'(n^{1/3})}$ for the universality problem for
an NFA. Note that a lower bound
for the universality problem implies the same lower bound
for the comparison problem: as taking $N_1$
to be the automaton accepting all inputs solves
the universality problem for $N_2$.
$3$-occur $3$SAT (SAT) is a variation of $3$SAT (SAT) in 
which each variable occurs at most three times.
Tan \cite{Ta22} provides an alternative proof for Fernau and Krebs result
mentioned above using a coding of $3$-occur $3$SAT. 
Best previously known upper bound for comparison problem was
by first converting NFAs into DFAs (Chrobak \cite{Ch86}
does this in time $2^{O((n \log n)^{1/2})}$) and then
doing the comparison.
The present work provides a faster algorithm which compares 
(with respect to inclusion) two unary $n$-state NFAs in time 
$2^{O(n \log n)^{1/3}}$.
This result is nearly optimal: within a factor 
$O((\log n)^{1/3})$ in the exponent,
given the above stated lower bound of $c^{n^{1/3}}$
by \cite{FK17}, for infinitely many inputs, under the assumption of ETH.

Recall that for an unambiguous nondeterministic finite automata (UFA),
every word outside the language has zero accepting runs and every
word inside the language has exactly one accepting run -- see below for more
explanations for these technical terms. Prior research
had established that the intersection of two $n$-state UFAs can be
represented by an $O(n^2)$-state UFA and that, over the binary alphabet,
the Kleene star of an $n$-state UFA can be 
represented
by an $O(n^2)$-state UFA \cite{Ce13,Co15,HK11,JO18,Pi00,PS02,YZS94}.
These bounds for intersection and Kleene star were optimal within
a constant factor. But the size-increase for the other regular operations
(complement, union, concatenation) remained open. It was however known
that disjoint union has linear complexity.

\begin{figure}[t]
\caption{Table of results on State Complexity. Here $c(n)=n^{\log n+O(1)}$.
The bounds for union and intersection of arbitrarily many $n$-state UFAs are
matching and reached when using $\Theta(n/\log n)$ many UFAs. Specific known
formulas for combining $k$ arbitrary UFAs were adjusted to this case.
Furthermore, let $LCM_n$ be the least common multiple of the natural
numbers from $1$ to $n$. $LCM_n \in 2^{\Theta(n)}$. Pointers to results
obtained by the authors in this paper are fully in blue.}
\label{tableresa}
\begin{center}
\begin{tabular}{|c|c|c|c|c|}
\hline
Operation & Lower Bound & Source & Upper Bound & Source \\
\hline
State Complexity & & & & \\
\hline
UFA Intersection & $n^2-n$ & Holzer and &  $n^2$ & Holzer and \\ 
   & & Kutrib \cite{HK02} & & Kutrib \cite{HK02} \\
Intersection & $LCM_n$ & 
   {\color{blue}Proposition~\ref{prop:regularoperationbounds}~(c)}
   & $n+LCM_n$ &
   {\color{blue}Proposition~\ref{prop:regularoperationbounds}~(k)} \\
   of $n$-state UFAs & in $2^{\Theta(n)}$ & &
	in $2^{\Theta(n)}$ & Holzer and \\
   & & & & Kutrib \cite{HK02} \\
UFA Complement   & $n^{(\log \log \log n)^{\Omega(1)}}$ & Raskin \cite{Ra18} & 
   $c(n)$ & {\color{blue} Theorem~\ref{th:complement}} \\
UFA Disjoint & $2n-4$ & -- & $2n$ & Jir\'askov\'a and \\
   Union & & & & Okhotin \cite{JO18} \\
UFA Union & $n(n-1)$ & Okhotin & $n + n \cdot c(n)$ &
   {\color{blue} Proposition~\ref{prop:regularoperationbounds}} \\
   & & \cite[Lemma 5]{Ok12} & & \\
   Union of & $LCM_n$ 
   & Proposition~\ref{prop:regularoperationbounds}~(g) &
   $n+LCM_n$ & 
   {\color{blue} Proposition~\ref{prop:regularoperationbounds}~(k)} \\
   $n$-state UFAs & in $2^{\Theta(n)}$
   & Okhotin \cite{Ok12}  & in $2^{\Theta(n)}$ &  \\
UFA Symm.\ diff.\ & $n^{(\log \log \log n)^{\Omega(1)}}$ &
   Raskin \cite{Ra18} & $2n \cdot c(n)$ &
   {\color{blue}Theorem~\ref{th:complement}} \\
Kleene Star & $(n-1)^2+1$ & Yu, Zhuang & $(n-1)^2+1$ & Yu, Zhuang \\
   & & and Salomaa \cite{YZS94} & & and Salomaa \cite{YZS94} \\
	UFA Concatenation & $2^{\Omega((n\log^2 n)^{1/3})}$ &
   {\color{blue}Theorem~\ref{th:concat}}
   & $2^{O((n \log^2 n)^{1/3})}$ & Okhotin \cite{Ok12}\\
\hline
\end{tabular}
\end{center}
\end{figure}

Unambiguous finite automata found much attention in recent
research, with a quasipolynomial lower bound 
$\Omega'(n^{(\log \log \log n)^q})$, for some positive rational constant $q$,
for the blow-up in the size for complementation by Raskin \cite{Ra18}.
Thus, it is impossible to achieve
a polynomial sized complementation of UFAs for the unary alphabet.
For comparison with the  binary alphabet case, it should be mentioned that
G\"o\"os, Kiefer and Yuan \cite{GKY22} constructed a
family of languages $L_n$ recognised
by an $n$-state UFA such that the size of the smallest NFAs
recognising the complement grows $n^{\Omega((\log n)/\polylog(\log n))}$.
This lower bound also trivially applies to UFAs.
However, the present work shows that, for the unary alphabet,
all regular operations
on UFAs can be done with at most quasipolynomial size-increase
except for concatenation.
This confirms a weak version of a conjecture of Colcombet \cite{Co15}
--- who predicted originally a polynomial size-increase for Boolean
operations.
For concatenation, the present work provides
the lower bound of $2^{\Omega((n\log^2 n)^{1/3})}$.
For larger alphabets, there is a big gap between the lower bound
$n^{\log n / \polylog(\log n)}$
by G\"o\"os, Kiefer and Yuan \cite{GKY22} and the upper bounds
for complementing an $n$-state UFA of
$2^{0.79n+\log n}$ by Jir\'asek, Jr., Jir\'askov\'a and  \v Sebej \cite{JJS18}
and $\sqrt{n+1} \cdot 2^{n/2}$ by Indzhev and Kiefer \cite{IK22}.

For unary automata, Okhotin \cite{Ok12} determined the 
worst-case size complexity for the determinisation of UFAs as 
$2^{\Theta((n \log^2(n))^{1/3})}$. This is thus also an upper bound
on the size-increase for complementation.
In his master's thesis, D\k ebski \cite{De17}
constructed an algorithm to complement unambiguous automata
over the unary alphabet while maintaining the upper size-bound of
$n^{O(\log n)}$ for these automata. The present work provides an alternative
construction with the slightly improved bound $n^{\log n + O(1)}$, so the
constant in the exponent is additive instead of multiplicative.

Note that, for the unary alphabet,
it is not efficient to compare two UFAs $U_1, U_2$ accepting
$L_1$ and $L_2$ respectively, with respect to inclusion,
by first constructing
a UFA for $\overline{L_2}$, and then a UFA for $L_1 \cap \overline{L_2}$ 
and then checking for emptyness.
Instead, direct comparison algorithms are used. Stearns and Hunt
\cite{SH85} provided a polynomial time algorithm for this. This paper
slightly improves it by providing a \textbf{LOGSPACE} algorithm
in the case that the UFAs are in the Chrobak Normal Form,
and an \textbf{NLOGSPACE} algorithm without any assumption
about the input UFAs being in the Chrobak Normal Form.
Note that the transformation of a UFA into the Chrobak Normal Form
can be done in polynomial time without increasing the number of states.
However, an \textbf{NLOGSPACE} algorithm for converting
a UFA into the Chrobak Normal Form might cause a polynomial size increase.

For unary alphabet, the $\omega$-word of a language $L$
is $L(0)L(1)L(2)\ldots$, where $L(k)$ is $1$ if the word of length
$k$ is in the language $L$ and $0$ otherwise.
This paper investigates the complexity
of deciding membership of the $\omega$-word of 
the language recognised by the given automaton $N$,
in some fixed $\omega$-regular language.
When, the automaton considered are DFAs,
this membership testing can be done in polynomial time. 
For UFAs and NFAs there is a known trivial upper bound
obtained by converting these automata into DFAs.
This upper bound has, up to logarithmic factors,
the exponent $n^{1/2}$ for NFAs and $n^{1/3}$ for UFAs.
The present paper gives lower bounds which match the above upper bound
up to a logarithmic factor in the exponent,
assuming the Exponential Time Hypothesis.

Tables~\ref{tableresa} and~\ref{tableresb} summarise
the results for UFAs.
Here $c(n) = n^{\log n + O(1)}$ and results
with a theorem/proposition plus number are proven in the present work. 
Table~\ref{tableresa} gives the
results on state complexity for the operations.
Table~\ref{tableresb} gives the space/time complexity bounds
for computing the corresponding UFAs.
Here, some of the lower bounds are assuming~\ETH.
Note that the lower bounds on size imply lower bounds on computations. 

\begin{figure}[t]
\caption{Table of results on Computational Complexity.
Here $c(n)=n^{\log n+O(1)}$.}
\label{tableresb}
\begin{center}
\begin{tabular}{|c|c|c|c|c|}
\hline
Operation & Lower Bound & Source & Upper Bound & Source \\
\hline
	Space Complexity & Lower Bound using ETH & & & \\
\hline
NFA Comparison & $\Omega'(n^{1/3})$ & Fernau and &
  $O((n \log n)^{1/3})$ & {\color{blue} Theorem~\ref{thm:algo}} \\
  & & Krebs \cite{FK17} & & \\
UFA Universality & -- & -- & $O(\log n)$ &
  {\color{blue}Theorem~\ref{th:logspace}} \\
in Chrobak NF & & & & \\
UFA Universality & -- & -- & $O(\log^2 n)$ &
  {\color{blue}Theorem~\ref{th:logspace}} \\
\hline
	Time Complexity & Lower Bound using ETH & & & \\
\hline
UFA Concatenation
	& $2^{\Omega((n \log^2 n)^{1/3})}$ & {\color{blue}Theorem~\ref{th:concat}}
   & $2^{O((n \log^2 n)^{1/3})}$ & Okhotin \cite{Ok12}\\
UFA Formulas & $2^{\Omega'((n\log n)^{1/3})}$ &
   {\color{blue}Theorem~\ref{thm:ethformula}} &
   $2^{O((n \log^2 n)^{1/3})}$ & Okhotin \cite{Ok12} \\
UFA-word in & $2^{\Omega'((n \log n)^{1/3})}$ &
   {\color{blue}Theorem~\ref{th:ufaomega}} & $2^{O((n \log^2 n)^{1/3})}$ & DFA-conversion \\
$\omega$-regular language & & & & \\
NFA-word in & $2^{\Omega'((n \log\log n/\log n)^{1/2})}$ & 
   {\color{blue} Theorem~\ref{th:nfaomega}} &
   $2^{O((n \log n)^{1/2})}$ & DFA-conversion \\
$\omega$-regular language & & & & \\
NFA Comparison & $2^{\Omega'(n^{1/3})}$ & Fernau and &
  $2^{O((n \log n)^{1/3})}$ & {\color{blue}Theorem~\ref{thm:algo}} \\
  & & Krebs \cite{FK17} & & \\
\hline
Time Complexity & No ETH Assumption & & & \\
\hline
UFA Universality & -- & -- & $O(n^{3/2} \log n)$ &
  {\color{blue}Theorem~\ref{th:logspace}} \\
in Chrobak NF & & & & \\ 
UFA Comparison & -- & -- & Poly(n) & Stearns and \\
   & & & & Hunt III \cite{SH85} \\
UFA Complement  & $n^{(\log \log \log n)^{\Omega(1)}}$ & Raskin \cite{Ra18} & 
   $n^{O(\log n)}$ & {\color{blue} Theorem~\ref{th:complement}} \\
UFA Formulas, & & & $n^{\log^{O(1)} n}$ &
  {\color{blue}Proposition~\ref{prop:regularoperationbounds}}\\
no Concatenation &&&& \\
\hline
\end{tabular}
\end{center}
\end{figure}

Note that in general for computing the complement, union, intersection and
for comparing NFAs, the computation space $O(f(n))$ corresponds
to the computation time $2^{O(f(n))}$ in the worst case.
Thus, every further improvement in the space usage will
also have an improvement in the time usage. 
UFA comparison is in \textbf{NLOGSPACE},
which is more or less optimal, and this corresponds to the polynomial time used
for the comparison.

\section{Details of Technical Concepts and Methods}\label{sec:detailsprelim}

\noindent
This section describes the following concepts which
are important for the current work.
\begin{itemize}
\item Finite automata are the main topic of the paper. This paper
deals with the special case of the unary alphabet and in general,
all results apply only to this special case.
\item Conditional lower bounds assume the Exponential Time
Hypothesis, an assumption stronger than $P \neq NP$,
which allows to prove good lower bounds. 
\item The prime number theorem is a useful tool for proofs.
Its main message is that below $n$ there are $\Theta(n/\log n)$ many
prime numbers.
\end{itemize}

\subsection{Finite Automata}

\noindent
A finite state automaton, see for example \cite{HMU07},
is a tuple $(Q,\Sigma,Q_0,\delta,F)$,
where $Q$ is a finite set of states, $\Sigma$ is a finite
alphabet (unary for this paper), $Q_0 \subseteq Q$ is a set of initial states,
$\delta$ is a transition function mapping $Q \times \Sigma$ to a subset of $Q$,
and $F \subseteq Q$ is a set of accepting states.
A run of the automaton on input $a_1a_2\ldots a_n$ is
a sequence of states $q_0, q_1,\ldots q_n$, such that
$q_0 \in Q_0$ and  for each $i<n$, $q_{i+1} \in \delta(q_i,a_{i+1})$.
The run is accepting if $q_n \in F$.
The input $a_1a_2\ldots a_n$ is accepted by the automaton if
there is a run on it which is accepting.
The set of words accepted by an automaton $A$ is called the
language recognised by the automaton (also sometimes called
the language of the automaton or the
language recognised by the automaton).
This language is denoted as $\Lang(A)$.
The terminologies word and string are used interchangeably.

Without postulating further constraints, the automaton is called
an \emph{nondeterministic finite automaton} (NFA).
An NFA which has, for each word in its language $L$,
exactly one accepting run is called an \emph{unambiguous
finite automaton} (UFA).
An NFA which has exactly one start state and for which
$\delta(q,a)$ has cardinality exactly one, for all
$q \in Q$ and $a \in \Sigma$,
is called a \emph{deterministic finite automaton} (DFA).
Note that every DFA has, for each word,
exactly one run and this run is
accepting iff the word is in the given language $L$.
Note that by the above definitions, every DFA is a UFA and
every UFA is an NFA.

A language $L$ is called \emph{regular} iff $L$ is the language
of some finite automaton, that is, some NFA. Note that every
regular language is also the language of a UFA and of a DFA,
the size of those might, however, be much larger than the
size of the smallest NFA.

In the case of the unary alphabet, every NFA or UFA can be brought into
a special form called the
Chrobak Normal Form \cite{Ch86}. For the unary alphabet,
an NFA is in the Chrobak Normal Form if and only if the
NFA consists of states $q_1,q_2,\ldots, q_s$,
$p_0^i, p_1^i,\ldots, p_{n_i-1}^i$, for $i$ with $1 \leq i \leq r$, for
some $s,r$ and $n_i$, such that the following conditions hold:
\begin{enumerate}
\item
If $s>0$, then $q_1$ is the starting state else
 $\{p_0^i: 1 \leq i \leq r\}$ is the set of starting states.
\item
$\delta(q_i,a)=\{q_{i+1}\}$, for $i$ with $1 \leq i<s$ and $a \in \Sigma$.
\item
If $s>0$ and $a \in \Sigma$, then $\delta(q_s,a)=\{p_0^i: 1 \leq i \leq r\}$.
\item
$\delta(p_j^i,a)=\{p_{j+1 \mod n_i}^i\}$, for $a \in \Sigma$,
$i \leq r$ and $j< n_i$.
\end{enumerate}

\medskip
\noindent
The states $q_1, q_2,\ldots, q_s$ and the transitions within
them is called the \emph{stem} of
the NFA, with $q_s$ the last state of the stem.
For $i$ with $1 \leq i \leq r$,
the states $(p_j^i)_{j<n_i}$ and the transitions among them,
are called the \emph{cycles} of the NFA, where $n_i$ is the
length of the cycle and $p_0^i$ is the \emph{entry state} of the cycle.
Which states of an NFA are accepting depends on its language.

When transforming an NFA into the Chrobak
Normal Form, the size might increase from $n$ to $O(n^2)$.
For example, consider the NFA with states $0,1,\ldots,n-1$, unary alphabet
and transitions $m \rightarrow m+1$, for $m=0,1,\ldots,n-2$,
and $m \rightarrow 0$, for $m=n-2,n-1$.
State $0$ is the starting state and the only accepting state.
Converting the above NFA to the Chrobak Normal Form
requires quadratic size increase. This Chrobak Normal Form NFA,
would consist of a stem of size $(n-2)(n-3)$,
followed by a single one-state-loop of an accepting state.
This example is typical, as the transformation of an NFA into Chrobak
Normal Form just increases the stem size and not the size of the cycle part.
However, in the case that the NFA is already a UFA, the size remains the
same \cite[Theorem 2.2]{JMR91}.

An important tool used is the following result of Okhotin \cite[Lemma 5]{Ok12},
which used prior work of Schmidt \cite{Sc78}.
Consider an NFA in the Chrobak Normal Form without a stem, where
for each cycle, all the states except the entry/start state
of the cycle are accepting. Suppose the lengths of the
cycles of the NFA are $p_1, p_2, \ldots, p_k$.
Note that the above NFA accepts all words of a length $\ell \neq 0 \mod p_i$
for some $i$. Let $L_{okh}$ be the corresponding language
and $p_1,p_2,\ldots,p_k$ its parameters.
Then, for any UFA recognising the language of the above NFA,
the number of states will be at least the LCM (least common
multiple) $p$ of $p_1, p_2, \ldots, p_k$. Furthermore,
there is indeed a one-cycle DFA of size $p$
recognising $L_{okh}$. Thus, the
lower bound matches the upper bound for the size of a UFA for
the language $L_{okh}$.
Technically, instead of the above LCM-bound, Okhotin \cite{Ok12} showed the
lower bound of $p+1$ for the union of $L_{okh}$ and the set of the
empty word. Okhotin's bound can be easily modified to the bound mentioned above,
by showing that any upper bound $q$ breaking the lower bound $p$ would
give an upper bound $q+1$ for Okhotin's language -- this can be obtained
by introducing a stem of length $1$ before the start node and shifting
the accepting states accordingly. Thus Okhotin's bound holds also
for $L_{okh}$. When used below, the language $H_{okh}$ denotes
the complement of $L_{okh}$.

For theorems below, when giving NFA or UFA in the Chrobak Normal Form as
input to an algorithm, it is assumed that first the stem is given and
then each of the cycles are given. Furthermore, the accepting states
are marked as such when the state names are given.

\subsection{The Exponential Time Hypothesis (ETH)}

\noindent
Impagliazzo, Paturi and Zane \cite{IP01,IPZ01} observed that for many
concepts similar to $3$SAT the running time of known algorithms
is exponential in the number $n$ of variables.
They investigated this topic and formulated what is
now known as the Exponential Time Hypothesis: There is a constant $c>1$
such that, every algorithm which solves $3$SAT, has for infinitely many
values of $n$ the worst case time complexity of at least $c^n$.
In order to handle lower bounds which occur only infinitely often,
the following definition is introduced:
$$
   f \in \Omega'(g(n)) \Leftrightarrow (\exists c>0)(\forall m)(\exists n>m)
     [f(n) \geq c \cdot g(n)],
$$
where the $m,n$ range over natural numbers and $f,g$ are functions with
positive values, that is, rational or real values strictly above $0$.
Then for many problems a conditional lower bound, that is, a lower bound
implied by the Exponential Time Hypothesis can be obtained.
For example, Fernau and Krebs \cite{FK17} showed for the
unary NFA universality problem the conditional lower
bound of $2^{\Omega'(n^{1/3})}$.
Note that, in the definition of $\Omega'$, one fixed constant $c$ works
for all the algorithms solving the problem.

In the present work, several
lower bounds for the runtime to solve problems related to unary UFAs
or NFAs are formulated under the assumption of the Exponential Time
Hypothesis using $\Omega'$-expressions.
Let $d$-occur $3$SAT denote the $3$SAT problem where each variable occurs 
at most $d$ times.
For constructing lower bounds, the following additional
result of Impaglazzio, Paturi and Zane \cite{IP01,IPZ01} is important.
Assuming the Exponential Time Hypothesis, there are constants $c,d > 1$
such that, for every algorithm solving $d$-occur $3$SAT,
there are infinitely many $n$ for which the run time of the algorithm 
is at least $c^n$ on some $n$-variable instance of $d$-occur $3$SAT.
Furthermore, $d$ can be chosen to be $3$,
at the expense of a possibly smaller
$c$, but still with $c>1$.

Furthermore, note that the Exponential Time Hypothesis comes in two
versions. Version (a) says that there is a constant
$c>1$ such that, for every algorithm solving $3$SAT, 
there are infinitely many $n$ for which the runtime
of the algorithm on some $n$-variable instance is 
at least $c^n$. Version (b) says that, for
every algorithm solving $3$SAT,
there is a constant $c>1$ such that, for infinitely
many $n$, the runtime of the algorithm
on some $n$-variable instance is at least $c^n$.
Version (b) is equivalent to saying that the
runtime is not in $2^{o(n)}$, version (a) is equivalent to saying that the
runtime is at least $2^{\Omega'(n)}$. The present paper uses version (a)
of the Exponential Time Hypothesis and version (a) implies version (b),
but not vice versa. Note that Impagliazzo and Paturi \cite{IP01}
use version (a) in their first paper.

\subsection{The Prime Number Theorem}

\noindent
The prime number theorem says that ratio of the $n$-th prime
number and
$n \cdot \log n / \log 2.71828\ldots$ converges to $1$
where $2.71828\ldots$ is the Euler's number.
A direct consequence of the prime number theorem is that
there is a constant $c$ such that, for each $n \geq 2$, there are
$n/\log n$ prime numbers between $n$ and $cn$.
In this paper,
variations of these consequences are used in constructions
to code $3$SAT instances into NFAs and UFAs in order to show the
hardness of decision problems.
Often in proofs, the constructed NFA is in the
Chrobak Normal Form
and consists of a stem of $s$ states, where $s=0$ is possible,
and disjoint cycles $C_1, C_2,\ldots$ of
lengths $p_1,p_2,\ldots$,
with $n=s+p_1+p_2+\ldots$ being the overall number of states.
In these cases, the states in cycle $C_i$ will be considered
to be ordered as $0$-th, $1$-st, $\ldots$ states in the cycle,
where the unary alphabet takes the NFA from the current to the
next state modulo $p_i$ in the cycle $C_i$. If $s>0$ the $0$-th state
in the cycle is the entry point into the cycle from the last state of
the stem else each cycle has a start state which is its $0$-th state.
Assume that the stem has $s$ states
and an input of length $t$ has been processed. If $s>t$, then
the NFA is in the $t$-th state of the stem else the NFA is in
the $(t-s \mod p_i)$-th state of the cycle $C_i$ for some $i$.

Often (though not always) the $p_i$ above would be either distinct
prime numbers or distinct prime numbers times a common factor,
the latter is in particular used for UFAs.
This allows us to use the Chinese Remainder Theorem to get that some possible
combination of states is reachable in the different cycles
for the same input word.
In the case that all cycle lengths
are pairwise coprime, the above holds
for every combination of states in the different cycles.

For example,
in Proposition~\ref{propone} below, cycles have a length which is
either a prime number or the product of two primes.
Every combination of states in the prime number length
cycles can be reached.
States in the cycles whose length are
the product of two prime numbers correspond 
to the ``coordinate-states'' in the corresponding
two cycles of prime number length.

\section{The Nondeterministic Finite Automata Comparison Algorithm}

\noindent
Fernau and Krebs \cite{FK17} showed a conditional (on ETH)
lower bound for comparison problem for NFAs. A proof is included
for the reader's convenience, as later results build on this method.
The result is by Fernau and Krebs \cite{FK17} and the proof used here is by
Tan \cite{Ta22}.

\begin{propo}[Fernau and Krebs \cite{FK17}, Tan \cite{Ta22}] \label{propone}
Given an $m$-variable $3$-occur $3$SAT instance, one can
construct in polynomial time an $n = \Theta(m^3)$ sized NFA
such that this NFA accepts
all words over the unary alphabet iff the given $3$-occur $3$SAT instance
is unsolvable. Thus, assuming that Exponential Time Hypothesis holds,
unary NFA universality requires $2^{\Omega'(n^{1/3})}$
computation time.
\end{propo}

\begin{proof}
Suppose an $m$-variable $3$SAT instance with at most $3m$ clauses, where
each variable occurs at most three times, is given. Without loss of generality assume
$m \geq 8$ and $\log m$ is a whole number. 
Let the clauses be $u_1, u_2, \ldots, u_{m'}$ (where $m' \leq 3m$)
and the variables be $x_1, \ldots, x_m$. 

Let $r=\lfloor{\frac{\log m}{3}} \rfloor$, and $s=\lceil{m'/r}\rceil$.
Consider the primes $p_0,p_1,\ldots, p_{s-1}$, where 
$8m \leq p_i \leq c'm$, for some constant $c'$. Note that by the prime
number theorem there exists such a constant $c'$.

Assign to each prime $p_i$, the clauses $u_{(i-1) \cdot r+1}, \ldots,
u_{i \cdot r}$.
Intuitively, for each $i<s$, there will be a cycle of length $p_i$ which 
will explore all possible truth assignments to variables in the clauses assigned
to $p_i$, and check whether they satisfy the corresponding clauses assigned.
Consistency of truth assignments to variables across clauses assigned to
different primes $p_i$ and $p_j$ will be checked using a cycle
of size $p_i  \cdot p_j$. Further details can be found below.

Note that $r$ clauses have at most $3r$ literals, and thus the number of
possible truth assignments to these literals is at most $2^{3r}$
with $2^{3r} \leq m \leq p_i$.
Order these assignments in some way so that one can say $k$-th truth
assignment, starting with $k=0$.

(1) For each $i<s$, form a cycle of length $p_i$. The $k$-th state (for $k<2^{3r}$)
in this cycle is rejecting iff the $k$-th truth assignment to the 
$3r$ literals in the clauses assigned to $p_i$ satisfy all the clauses assigned
	to $p_i$. Note that if $k\geq 2^{3r}$, then the $k$-th state is accepting.

(2) For each pair $i,j$ such that the clauses assigned to
$p_i$ and $p_j$ have a common variable, form a cycle of length $p_i \cdot p_j$.
The $k$-th state in this cycle is accepting iff 
the $(k \mod p_i)$-th truth assignment to the literals in the clauses 
assigned to $p_i$
and the $(k \mod p_j)$-th truth assignment to the literals in the clauses 
assigned to $p_j$ are inconsistent within or with each other.

The starting states of the above NFA are the $0$-th state in each
of the cycles.
Now note that the above NFA rejects the unary word of length $\ell$ iff 
the following are satisfied:

(A) For each $i<s$, the $(\ell \mod p_i)$-th truth assignment to the literals
in clauses assigned to $p_i$ satisfy the clauses assigned to $p_i$.

(B) For each $i,j<s$, if clauses assigned to $p_i$ and $p_j$ have 
a common variable, then the $(\ell\mod p_i)$-th and the $(\ell \mod p_j)$-th
truth assignment to the literals in clauses assigned to $p_i$ and $p_j$
respectively are consistent.

Thus, the language recognised by the above NFA is universal iff the $3$SAT
formula is not satisfiable.

The number of states in the above NFA is bounded by  $(3m/r) \cdot c'm$ (for
cycles in (1)), plus $(c'm)^2 \cdot 3m$ (for cycles in (2), as there are
$m$ variables each appearing at most thrice, so one needs to check at
most $3m$ pairs). Hence, the number of states is proportional to $m^3$.
It follows from the \ETH\ that the complexity of testing universality
for $n$-state NFA is at least $2^{\Omega'(n^{1/3})}$. 
\end{proof}

\noindent
The upper bound in the next result is only slightly larger than the
lower bound given above. More
precisely, the exponent of the upper bound is 
above the exponent $n^{1/3}$ of the lower bound
by a multiplicative factor of $(\log n)^{1/3}$.
As the time used by an algorithm is at most $2^{O(\text{space used})}$,
any improvement in the space bound
given in the algorithm below would also result in the improvement
in the time bound.

\begin{theorem} \label{thm:algo}
Suppose two nondeterministic finite automata $N_1, N_2$
over the unary alphabet 
are given. Let $n$ denote the maximum of their number of states. 
Then, one can decide
whether $\Lang(N_1) \subseteq \Lang(N_2)$
in deterministic
time $O(c^{(n \log n)^{1/3}})$, for a suitable constant $c > 1$.
Furthermore, the algorithm can be adjusted such that
the space used is $O((n \log n)^{1/3})$.

As equality and universality can be checked using
comparison algorithm, the above time bound also applies for checking
equality of languages recognised by the two NFAs and for
checking universality of the language recognised by an NFA.
\end{theorem}

\begin{proof}
Without loss of generality assume $n$ is large enough so
that the prime number theorem and other bounds needed below apply.
In polynomial time in the number
of states $n$, a nondeterministic
finite automata can be transformed
into the Chrobak Normal Form \cite{Ch86},
where it consists of a stem
of up to $n^2$ states, followed by parallel cycles which, together, use
up to $n$ states. 
Let $N_1'$ and $N_2'$ denote the corresponding coverted
NFAs. Without loss of generality, assume that the stems of
$N_1'$ and $N_2'$ have the same length.
To see that this is without loss of generality,
note that in the Chrobak Normal Form, 
a stem can be made 
one longer by adding one state at the end of the stem 
and shifting the entry 
point into each cycle by one state.
The new state in the stem would be accepting iff one of the 
prior entry points in the cycles was accepting.
This can be done repeatedly (at most $O(n^2)$ times)
until the stems have the same length.

The comparison of the behaviour of the NFAs on the stems of equal length before
entering the cycles can be done by just checking if
the corresponding states
at the same distance from the start are both accepting or
both rejecting.
For comparing two NFAs in Chrobak Normal Form, the comparison
of the cycle part is therefore the difficult part.
Thus, for the following, assume without loss
of generality that $N_1'$ and $N_2'$ are in the Chrobak Normal
Form, and do not have any stem. These NFAs thus consist
only of disjoint cycles, each having one start state 
and the only nondeterminism is the choice of the
start state, that is, the cycle to be used.

Note that a cycle $C$ of length $m$ in an NFA in Chrobak Normal Form
can be converted
into a cycle $C'$ of length $w$, where $m$ divides $w$,
by having the $s$-th state of $C'$ as accepting if
$(s \mod m)$-th state of $C$ is accepting. This way, for
an appropriate value of $w$, one can combine several cycles,
whose lengths divide $w$, into one cycle of length $w$.
Suppose there is a set $X$ of a small number of $w$'s
(which pairwise have the same greatest common divisor (gcd)
$r$), such that the lengths
of all the cycles of $N_1'$ and $N_2'$ divide at least one
$w \in X$. Then converting the two NFAs $N_1'$ and $N_2'$ 
into $N_1''$ and $N_2''$ respectively, such that
$N_1''$ and $N_2''$ have
cycles only of lengths $w \in X$, 
allows for easier comparison.
The following argues for existance of such $X$, and constructs
the corresponding $N_1''$ and $N_2''$.
These converted NFAs $N_1''$ and $N_2''$ 
are called comparison normal form NFAs below.

Intuitively, lengths of only few
(at most $(n/ \log^2 n)^{1/3}$) cycles of $N_1'$ and $N_2'$ can have two large
(at least $(n \log n)^{1/3}$) prime factors. The aim
of $P$ defined below is to collect all such prime factors along with the
small primes.
Set $Q$ defined below collects the large primes
which are not in $P$.
Thus, $P$ and $Q$ together provide all the prime factors
of the lengths of the cycles of the two NFAs $N_1'$ and $N_2'$.

Let $P=\{$prime number $p:
p < (n \log n)^{1/3}$ or there exist a prime
$q \geq (n \log n)^{1/3}$ such that at least one of the
NFAs $N_1'$ and $N_2'$ has a cycle with length divisible by 
$p \cdot q \}$.

Note that the number of primes smaller than $(n \log n)^{1/3}$
is at most $O((n/\log^2 n)^{1/3})$. 
Note that the cycles
within each NFA $N_1'$ or $N_2'$ are disjoint and sum of their lengths
is bounded by $n$. Thus,
the number of primes $p \geq (n\log n)^{1/3}$ such that
the length of some cycle in one of $N_1'$ and $N_2'$ is divisible by
$p \cdot q$ for some prime $q \geq (n\log n)^{1/3}$,
is at most $O((n/\log^2 n)^{1/3})$.
It follows that the cardinality of $P$ is at most $O((n/\log^2 n)^{1/3})$.

Let $Q=\{$prime number $p \leq n : p \not \in P\} \cup \{1\}$.

For each $p \in P$, let $k_p$ be maximum number $k$ such that
$p^k \leq n$. Note that $p^{k_p}$ is the highest
power of $p \in P$
which could divide the length of some cycle in
$N_1'$ or $N_2'$. Similarly, $q^2$ is the highest power
of $q \in Q-\{1\}$ which could divide the length of any cycle in
$N_1'$ or $N_2'$.
Let $r$ be the product of all $p^{k_p}$, $p \in P$.
Note that $r$ is in $O(n^{c' \cdot (n / \log^2 n)^{1/3}})
= O(2^{c' \cdot (n \log n)^{1/3}})$, for
some constant $c'$.
Thus, $r \leq c^{(n \log n)^{1/3}}$ for some constant $c > 1$.

Let $X=\{r\cdot q^2: q \in Q\}$.

Note that the lengths of all the cycles in $N_1'$ or $N_2'$
divide some $w \in X$. Moreover, the gcd of any two numbers
in $X$ is $r$.
Now, for the ease of comparing $N_1'$ and $N_2'$,
one transforms each of these NFAs into an equivalent 
``comparison normal form'' NFA of size at most 
$r \cdot n^3$ as follows.

For each $q \in Q$, the comparison normal form NFA
$N_1''$ (respectively $N_2''$) has a cycle of length $r \cdot q^2$.
The $s$-th state in this cycle of length $r \cdot q^2$
is accepting iff there is a cycle
of length $p$ in $N_1'$ (respectively $N_2'$), where $p$ divides $r \cdot q^2$
and $s \mod p$-th state in this cycle
is accepting. Note that the comparison normal
form NFAs $N_1''$ and $N_2''$ accept the same language 
as $N_1'$ and $N_2'$ respectively.
The comparison normal form NFAs $N_1''$ and $N_2''$
can be constructed in time $r \cdot Poly(n)$ by constructing
each cycle separately, and determining its
accepting states by considering all cycles
of length $p$ in $N_1'$ and $N_2'$ respectively,
where $p$ divides $r \cdot q^2$.

As $N_1''$ and $N_2''$ accept the same languages as $N_1'$ and $N_2'$
respectively,
it suffices to compare $N_1''$ and $N_2'"$.

Now $\Lang(N_1'') \subseteq \Lang(N_2'')$ 
iff for all $s < r$ one of the following two
options holds:

(A) There is a $q \in Q$ such that in $N_2''$,
for all $t<q^2$, the $(s + t \cdot r)$-th state in the cycle of length
$r \cdot q^2$ is accepting.

(B) For every $q \in Q$ and for all $t<q^2$, 
if the $(s + t \cdot r)$-th state,
in the cycle of length $r \cdot q^2$ is accepting in
$N_1''$, then it is also accepting in the
corresponding cycle of
$N_2''$.

This condition can be checked in time $r \cdot Poly(n)$: There are $r$
possible values of $s$ and for each such $s$, check needs to be
done only for $O(n^3)$ states, namely for each $q \in Q$,
the $(s+t \cdot r)$-th state, where $t \in \{0,1,\ldots,q^2-1\}$. 
Note that $q^2 \leq n^2$.

For correctness, it is first shown that (A) and (B) are sufficient
conditions. Let $s<r$ be given. 

If (A) is satisfied, then $N_2''$
accepts, for all $t$, all words of length $s+t \cdot r$,
as for the given $q$, all these words are accepted by the
cycle of length $r \cdot q^2$ in $N_2''$.

Now suppose (B) is satisfied. Consider a string
of length $s+t \cdot r$, (for some $t$) accepted by $N_1''$.
Thus, there is a $q \in Q$ such that, in the cycle of length $r \cdot q^2$
in $N_1''$,
the $(s+t \cdot r) \mod (r \cdot q^2)$-th state is accepting for $N_1''$.
From the statement of condition (B) it follows that in $N_2''$, in
the corresponding cycle of length $r \cdot q^2$,
the $(s+t \cdot r) \mod (r \cdot q^2)$-th state is accepting.
Therefore $N_2''$ also accepts the string of length $s+t \cdot r$.
It follows that $N_2''$ accepts all strings of length
$s \mod r$ which are accepted by $N_1''$.

As one of (A) or (B) holds for each $s<r$, it follows that the
language recognised by $N_1''$ is contained
in the language recognised by $N_2''$.

For the converse, assume that the following
condition (C) holds for some $s<r$:
For every $q \in Q$ there exists a $t_q$ such that:
(i) the
$(s + t_q \cdot r)$-th state is rejecting in the cycle of length $r \cdot q^2$
in $N_2''$ and furthermore, (ii) for at least one $q$, the 
$(s + t_q \cdot r)$-th state is accepting in the cycle of length
$r \cdot q^2$ in $N_1''$. So (C) is true iff both
(A) and (B) are false. 
Now, by the Chinese Remainder Theorem, there exist an $s'$ such that
$s' \mod r\cdot q^2= s+t_q\cdot r$, for each $q \in Q$.
Thus, $N_1''$ accepts the unary string of length
$s'$ (as, for at least one $q \in Q$, it has a cycle of length $r \cdot q^2$
in which $(s+t_q \cdot r)$-th state is accepting),
while $N_2''$ does not accept the unary string of
length $s'$ (as, for all $q \in Q$,
$(s+t_q \cdot r)$-th state in the cycle of
length $r \cdot q^2$ in $N_2''$ is non-accepting).

For the space-bounded variant of the algorithm, the algorithm cannot bring
the automaton into a normal form, as that cannot be stored
within the space allowed. The comparison algorithm
therefore has the translation into the above used normal form more implicit.
Note that one can generate the Chrobak Normal Form of
an NFA in \textbf{NLOGSPACE} using the algorithm given by
Chrobak \cite{Ch86}, and thus can obtain the needed
information about the cycle sizes, accepting states etc
in the following proof as needed.

Now note the following about the above time bounded comparison algorithm.
$\Lang(N_1'') \subseteq \Lang(N_2'')$ iff
$(\forall q \in Q) (\forall s < r) (\forall m < q^2)
[$ If $s+m\cdot r$-th state in the cycle of length
    $r\cdot q^2$ in $N_1''$ is accepting, then 
$[$ $s+m\cdot r$-th state in the cycle of length $r\cdot q^2$ in $N_2''$ is accepting $]$
or $(\exists q' \in Q)(\forall \ell <(q')^2)[$ $s+\ell\cdot r$-th state in the cycle of length 
    $r\cdot (q')^2$ in $N_2''$ is accepting $]$
 $]$.

In the above, when $q \neq q'$, the requirement:
$(\forall \ell<(q')^2)[$ $s+\ell\cdot r$-th state in the cycle of length 
$r\cdot (q')^2$ in $N_2''$ is accepting $]$
is equivalent to saying:
$(\forall \ell<n^2)[$ $(s+m \cdot r+\ell\cdot r q^2) \mod (r\cdot (q')^2)$-th state
in the cycle of length $r\cdot (q')^2$ in $N_2''$ is accepting $]$.

Thus,
$\Lang(N_1'') \subseteq \Lang(N_2'')$ iff
$(\forall q \in Q) (\forall s < r\cdot n^2)
[$ If the string of length $s$ is accepted by the cycle of length
    $r\cdot q^2$ in $N_1''$, then $(\exists q' \in Q)(\forall \ell <n^2)
[$ the string of length $s+\ell rq^2$ is
accepted by the cycle of length $r \cdot (q')^2$ in $N_2'']]$.
This is what is implemented by the procedure below.

The following needs constantly many variables bounded
by $2n^4 \cdot r$, these variables can be stored
in $O(\log r) = O((n \log n)^{1/3})$ space. 
Note that $s$-th state in cycle of size $r \cdot q^2$ being
accepting in $N_1''$ (respectively $N_2''$) is
equivalent to there being a cycle of size $p$ dividing $r \cdot q^2$
in $N_1'$ (respectively $N_2'$) such that
$s \mod p$-th state in this cycle is accepting.

First compute $r$. 
Second, for all cycles $C$ in $N_1'$ and all $s < n^2 \cdot r$
do the following check.
Suppose $C$ is of length $p$, and $s \mod p$-th state in $C$ is accepting.
If $p$ divides $r$, then let $q=1$ else
let $q$ be the unique prime such that $p$ divides $r q^2$. Note that $q \leq n$.
Now check if there is a number $q'$ such that either $q'=1$ or
$q'$ is a prime $\leq n$ which does not divide $r$, and for each
$\ell = 0,1,\ldots,n^2-1$ there is a cycle of
a length dividing $r q'^2$ in $N_2'$ 
which accepts the string of length
$s+\ell r q^2$.

Now $\Lang(N_1') \subseteq \Lang(N_2')$ (and thus $\Lang(N_1) 
\subseteq \Lang(N_2)$) iff all the above tests in
the algorithm have a positive answer. Note that $r$ is chosen such that
$\log c^{(n \log n)^{1/3}}$ and $\log r$ have the same order of magnitude
and thus $O(\log c^{(n \log n)^{1/3}}) = O(\log(r \cdot n^4))$. Therefore,
a real improvement of the space usage would also give an improvement
of the computation time.
\end{proof}

\section{Unambiguous Finite Automata and their Algorithmic Properties}

\noindent
Recall that an unambiguous automaton (UFA) satisfies that
for every input word, there
is either exactly one accepting run or none. On one hand, these are
more complicated to handle than nondeterministic finite automata so
that the union of $n$ $n$-state automata cannot be done with $n^2$ states.
On the other hand, they still, at least for the unary alphabets, have good
algorithmic properties with respect to regular operations (union, intersection,
complementation, Kleene star) and comparison
(subset and equality).

\subsection{Complementation}

\noindent
The following theorem shows that given an $n$-state UFA, a quasipolynomial 
in $n$ number of states and time is enough to construct a further UFA recognising
the complement of the language of the given UFA.

\begin{theorem} \label{th:complement}
Given a UFA $U$ with $n$ states, there is
another UFA $U'$ with $n^{\log(n)+O(1)}$ states
which recognises the complement of Lang$(U)$.
Furthermore, $U'$ can be computed in time
$n^{O(\log(n))}$.
\end{theorem}

\begin{proof}
Assume without loss of generality that $U$ is in
the Chrobak Normal Form. Furthermore, as inverting the states on the
stem is trivial, for easier notation,
it is assumed without loss of generality that
$U$ does not contain any stem and
consists only of $m$ disjoint cycles
$C_0,C_1,\ldots,C_{m-1}$ for some $m$,
each having exactly one start state 
(the $0$-th state of the cycle).

Intuitively, the idea is to output a UFA
using a recursive algorithm. At the start of
the recursion, the aim is to output a UFA
(without any stem)
which accepts exactly the strings 
having lengths $k \mod d$, with $k=0, d=1$,
which belong to the complement
of the language recognised by $U$.

At each step of the recursion, with parameters
$k,d$, either the algorithm 

(a) returns a UFA (without any stem) which accepts
exactly the strings in the complement which
have lengths $k \mod d$ or

(b) makes recursive calls to obtain UFAs (without any stem)
for accepting exactly the strings 
in the complement with lengths
$(k+d\cdot s) \mod (d \cdot \ell)$, 
for some value of $\ell$ and
$s$ being $0,1,\ldots, \ell-1$. As the languages recognised
by the above UFAs (which are without any stem) are pairwise disjoint
languages, the union of
these UFAs will give a UFA for
accepting exactly the strings in the complement which
have lengths $k \mod d$.

Though, the algorithm is presented as a recursive algorithm,
one can also view the solution as a tree, where
the root of the tree has parameters $(k=0,d=1)$. 
Any node of the tree is either a leaf (i.e., it 
gives a UFA, without any stem, for accepting exactly the 
strings in the complement which have lengths $k \mod d$), or has 
$\ell$ children, for some $\ell$, with parameters
$(k+sd, d \cdot \ell)$, for $s$ being $0,1,\ldots, \ell-1$
respectively in the $\ell$ children.
The UFA for the complement will thus be the union of the UFAs
at the leaves.
Note that for any two leaves with parameters $k',d'$ and
$k'',d''$ there is no length $m$ with $m$ being
same as both $k' \mod d'$ and $k'' \mod d''$.
Thus, the above tree is also called a tree
of different modulo residua.

Now the formal recursive algorithm is presented.
Initially the algorithm is called
with parameters $k=0$ and $d=1$. Thus,
the output UFA $U'$ is the output
of UFAcomplement$(0,1)$.

\medskip
\noindent Function UFAcomplement$(k,d)$
\begin{enumerate}
\item If there is a cycle $C_i$ in $U$ such that all 
      strings of
      length $k + sd$ with $s < |C_i|$ are accepted by this cycle,
      then return to the calling instance of the recursion 
      a UFA for emptyset (as
      $U$ accepts all strings of length $k \mod d$, UFA
      for the complement needs to reject all strings of 
      length $k \mod d$).
\item If there is no cycle $C_j$ accepting any string of
      length $k+sd$ with $s < |C_j|$, then return one cycle of 
      length $d$
      for which the $k$-th state is accepting and all other states
      are rejecting (as $U$ rejects all strings
      of length $k \mod d$, the UFA for complement needs to 
      accept all such strings).
\item Otherwise, there is a cycle $C_h$ which accepts some but not
      all strings whose length modulo $d$ is $k$. The algorithm computes now
      the least common multiple $d' = lcm(d,|C_h|)$ and makes,
      for $s=0,1,\ldots,d'/d-1$, a recursive call with the
      parameters $(k+ds,d')$. 

      Return the union of the answers obtained from
      the recursive calls.
\end{enumerate}
\noindent End of function UFAcomplement$(k,d)$

\medskip
\noindent
The algorithm clearly terminates, as when $d$ is the multiple of
all the cycle lengths and $k$ is a number 
between $0$ and $d-1$, then every
cycle $C_i$ has the property that it either accepts all strings
of length $k+sd$ or rejects all strings of length $k+sd$.
However, the following claim shows that the value of $d$ is much
smaller at the termination step.

\begin{claim}
The value of $d$ at the termination step is
at most $n\cdot (n/2) \cdot (n/2^2) \cdot \ldots 
\leq n^{0.5 \log n+c}$, for some constant $c$.
\end{claim}

\noindent
To see the claim,
consider any branch of the recursive descent, with the values
of $(k,d)$ in the recursive calls being 
$(k_0,d_0)$ (at the root), $(k_1,d_1)$, $\ldots$.
Suppose, in this branch, cycle $C_{e_i}$ is chosen in step 3,
when the values of $(k,d)$ were $(k_i,d_i)$ (except at the last level which 
terminates in step 1 or step 2). 
For $i$ not being the last level of the recursive descent,
the following properties hold:
\begin{enumerate}[(i)]
\item $d_{i+1}=lcm(d_i,|C_{e_i}|)$.
In particular, $d_i$ divides $d_{i+1}$.

\item $k_{i+1}=k_i+s_id_i$, for some $s_i$.

\item $C_{e_i}$ accepts some but not all strings of length
    $k_i \mod d_i$.

\item $|C_{e_i}|$ does not divide $d_i$ but divides $d_{i+1}$.
\end{enumerate}

\medskip
\noindent
Thus, for any levels $g,h$ not being the last level of
the recursive descent with $g<h$, using (i) and (ii) repeatedly,

$k_{h}=k_g+s_{g}d_{g}+\ldots+s_{h-1}d_{h-1}=k_g+s_h'd_g$,
for some $s_h'$, and $d_g$ divides $d_h$.

Thus, using (iii) both $C_{e_g}$ and $C_{e_h}$ accept some words of
length $k_g \mod d_g$. 
Thus,
there is a common factor $b>1$ of $|C_{e_g}|$ and $|C_{e_h}|$ which
does not divide $d_g$ (otherwise, $U$ will not be unambiguous).
Note that $b$ divides
$d_{g+1}$ as $|C_{e_g}|$ divides $d_{g+1}$ (by (iv)).
Thus, there is an extra common factor greater than $1$
between $d_{g+1}$ and $|C_{e_h}|$
compared to $d_g$ and $|C_{e_h}|$, for each $g<h$. 
Thus, common factor between
$d_g$ and $|C_{e_h}|$ is at least $2^g$. It follows that $d_{h+1}/d_h$
is at most $n/2^h$.

Thus, the number of levels is at most $\log n$ and the value of
$d$ at the termination step is at most $n\cdot (n/2) \cdot 
(n/2^2) \ldots \leq n^{0.5 \log n+c}$, 
for some constant $c$, where it can be safely assumed that $c =2$. This
proves the claim.

Also, it is easy to see by induction that
the automaton generated by the algorithm is a UFA, as
UFAcomplement$(k,d)$ either returns a UFA in steps 1 or 2,
or combines the UFAs generated by the recursive calls in step 3,
sets accepted by which are disjoint as they only accept strings
of length $k+ds \mod d'$, for different values of $s$.

The output automaton $U'$ is the union of cycles of length 
up to $n^{(\log n)/2+c}$.
In post-processing, one can unify distinct cycles of the
same length, $d$, with $k_1$-th, $\ldots$
$k_s$-th states as accepting into a single cycle of length
$d$ which has the $k_1$-th, $\ldots$, $k_s$-th states as accepting
states.
After this post-processing, there are at most $n^{(\log n)/2+c}$ cycles
and thus the overall size of the output UFA $U'$
is at most $n^{\log n+2c}$.

For the time bound, note that $U'$ can be detemined in time polynomial in
the number and size of the cycles of $U'$ and $U$. 
The theorem follows.
\end{proof}

\subsection{Comparison with Respect to Containment}

\noindent
The following results establish how to check whether languages
defined by UFAs have a subset relation or are equal or are
incomparable.

\begin{theorem} \label{th:logspace}
(a) Whether a UFA in the Chrobak Normal Form accepts all words
can be decided in \textbf{LOGSPACE}. 
The time complexity of the algorithm is $O(n^{3/2}\log n)$.

(b) Given UFAs $U_1$ and $U_2$ in the Chrobak Normal Form as input,
it can be decided in \textbf{LOGSPACE} whether $\Lang(U_1)
\subseteq \Lang(U_2)$.
\end{theorem}

\begin{proof}
(a) In the given UFA, first check if the states of the stem are all
accepting, which can clearly be done in \textbf{LOGSPACE},
as $O(\log n)$ memory is enough to track positions in the UFA.

Now, suppose the $k$-th cycle has $i_k$ accepting states and 
length $j_k$. Then, the UFA accepts all words, that is, is universal
iff $\sum_k i_k/j_k = 1$ (as each word is accepted
in one and only one cycle).
The following proof first gives an algorithm on how to check this
without being careful about time and space.
Later the algorithm is modified
to do it in the required space and time bounds.
For ease of notation, assume that all the cycle lengths are different.
(as same length cycles can be combined). This is done only for
the analysis of the algorithm without the space/time bound.
This assumption is not needed for the space/time bounded
algorithm given later.

As the computation with rational numbers might be prone to rounding, one
first normalises to one common denominator, namely $p = \prod_k j_k$ and
furthermore computes $s = \sum_k i_k \cdot \prod_{h \neq k} j_h$. Now the
above equality $\sum_k i_k/j_k = 1$ holds iff $s = p$.

The values of $s$ and $p$ can be computed iteratively by the following
algorithm. Note that there are at most $n$ cycles and each time a cycle
is processed in the input UFA,
the corresponding values $i_k$ and $j_k$ can be established. So the algorithm 
is as follows:
\begin{enumerate}
\item Initialise $s=0$ and $p=1$.
\item For each $k$ do Begin
	\begin{quotation}
        \noindent Find $i_k$ and $j_k$ of the corresponding cycle. \\
        Update $s = (s \cdot j_k)+(i_k \cdot p)$ and $p = p \cdot j_k$.
	\end{quotation}
       End (For).
\item If $s = p$, then accept else reject.
\end{enumerate}

\medskip
\noindent
In this algorithm, only the variables $p$ and $s$ need more space
than $O(\log n)$. The other variables all have value
between $0$ and $n$ which can be stored in $O(\log n)$ space.

As the sum of the lengths of all the cycles is
bounded by $n$, there can be at most $n^{1/2}$
cycles of length at least $n^{1/2}$.
Furthermore, there are at most $n^{1/2}$ cycles
with length shorter than $n^{1/2}$, due to different cycles
having different length. It follows that in total
there can be at most $2n^{1/2}$ cycles. 
As each cycle has length at most $n$,
it follows that the values of $s$ and $p$ are bounded by $n^{ 2(n^{1/2})}$.

By the prime number theorem, there exist $5 \cdot n^{1/2}+2$
primes $\leq O(n^{1/2} \log n)$.
The product of these primes is
larger than the upper bound $n^{1+2(n^{1/2})}$ of $s$
and $p$. Thus, using the Chinese Remainder Theorem,
$s = p$ iff $s\mod q = p \mod q$ for the first
$5 \cdot n^{1/2}+2$ primes $q$.
Thus, instead of using the above algorithm with exact numbers, one can
do the computation modulo the first $5 \cdot n^{1/2}+2$ primes.
The modified algorithm would be as follows.
\begin{enumerate}
\item Let $q=2$ and $\ell = 1$.
\item Initialise $s=0$ and $p=1$ (both are kept modulo $q$)
\item For each $k$ do Begin 
      \begin{quotation}
      \noindent Find $i_k$ and $j_k$ of the corresponding cycle. \\
      Update (both computations modulo $q$)
      $s = (s \cdot j_k)+(i_k \cdot p)$ and $p = p \cdot j_k$.
      \end{quotation}
      End (For).
\item If $s \neq p$ (modulo $q$), then reject.
\item Let $\ell = \ell+1$ and replace $q$ by the smallest prime above $q$.
      Note that the primality test for a number $q\in O(n^{1/2}\log n)$ can be done
      by checking divisibility by numbers below $\sqrt{q}$.
\item If $\ell \cdot \ell <35n+5$, then goto step 2 else accept.
\end{enumerate}

\medskip
\noindent
The condition $\ell \cdot \ell < 35n+5$ always applies when
$\ell \leq 5\sqrt{n}+2$.
For space usage, note that the primes $q$ and the variables $i_k, j_k, \ell$ 
used are all bounded by $n$. Thus, the values of
$s$ and $p$ modulo $q$ are also bounded by $n$. 
Primality tests can be done in $O(\log n)$ space for the 
usual way of doing it -- checking
all divisors up to the square root of the number.
Thus, the space needed for the above algorithm is $O(\log n)$.

Now consider the time bound for this 
algorithm.
Note that for each value of $q$, the algorithm transverses each of the cycles
of the input UFA to determine the corresponding $i_k$ and $j_k$. The cycles
are disjoint and have together at most $n$ states. For each of the cycles,
the multiplications in step 3 and taking $\mod$ in steps 3 and 4 take 
$O(\log n)$ time. Thus, the total time taken by steps 3 and 4, for
each value of $q$ is bounded by $O(n \log n)$. 
Primality testing of $q$ can also be done in $O(n \log n)$ time.
Thus, the algorithm takes $O(n \log n)$ time for each $q$.
Thus, the total time taken by the algorithm is bounded
by $O(n^{3/2} \log n)$. 

(b) Note that basically the same idea as in (a) can 
be used to check
if a UFA accepts all unary strings in sets $v w^*$ for some $v,w$ of length
up to $2n$.

Partition the strings accepted by $U_1$ into two
groups: 
\begin{enumerate}[(i)]
\item A finite set $X_1$ of strings of length at most $n$
(where $n$ is the size of the UFA) and
\item A set $X_2$ consisting of subsets of the form $vw^*$,
where $v,w$ are unary strings
with $n < |v| \leq 2n$ and $|w| \leq n$.
\end{enumerate}

\medskip
\noindent
The strings in group (ii) above are from the cycles in $U_1$: 
for each accepting state in a cycle in $U_1$,
pick $|w|$ as the length of the cycle and
$v$ as the smallest string of length $>n$ which
leads $U_1$ to this accepting state.

Whether strings in group (i) are accepted 
by $U_2$ can be easily checked, where if there is a 
branching into the cycles, one can do a depth first search.

For strings in group (ii), each set of the form $vw^*$, where $n < |v|  \leq 2n$
and $|w| \leq n$, is checked separately. As $|v|>$ the length of the 
stem part of $U_2$, one can first modify the cycle part 
of $U_2$ to always start in a state 
which is reached after $|v|$ steps, and ignore the stem part.
This would basically mean that one needs to check if
all words in $w^*$ are accepted in the
modified $U_2$ (here $U_2'$ denotes this modification of $U_2$).
For space constraints, note that one does not need to write down $U_2'$,
but just need to know the length by which the starting state of each cycle
is shifted (which is the difference between $|v|$ and the length of
the stem part of $U_2$).
Now, for checking whether every word in $w^*$ is accepted by $U_2'$,
consider a further modified $U_2''$ formed as follows:
for each cycle $C$ in $U_2'$
with length $r$ and states $s_0,s_1,\ldots,s_{r-1}$
($s_0$ being starting state, and transitions on unary
input being from $s_i$ to $s_{i+1}$, where $i+1$ is taken mod $r$) 
form a cycle $C'$ in $U_2''$ with states
$s'_0,s'_1,\ldots,s'_{r-1}$
($s'_0$ being starting state, and transitions on unary
input being from $s'_i$ to $s'_{i+1}$, where $i+1$ is taken mod $r$)
where $s'_i$ is an accepting state iff $s_{i\cdot |w| \mod r}$
was an accepting state in $C$.
This new UFA $U_2''$ also has at most $n$ states,
has the same number of cycles as $U_2'$,
with the length of cycles being
the same as the length of the corresponding cycles
in $U_2'$.
Now, similar to part (a), one just needs to check if $U_2''$
is accepting all unary strings.
Here again note that one doesn't need to write down $U_2''$ fully,
but just needs to check, for each cycle, its length and the number of
accepting states, which can be done in \textbf{LOGSPACE}.
\end{proof}

\noindent
As converting a UFA into the Chrobak Normal Form can be done in polynomial time
without increasing the size
and as logarithmic space computations are in polynomial time, 
one directly gets the following corollary.

\begin{corollary}[Stearns and Hunt \cite{SH85}] \label{co:logspace}
The universality problem and the inclusion problem for two
$n$-state UFAs can be decided in polynomial time.
\end{corollary}

\noindent
Corollary~\ref{co:logspace} can be improved to
computations in \textbf{NLOGSPACE}
and, by Savitch's Theorem \cite{Sa70}, in \textbf{DSPACE}$((\log n)^2)$.
This holds as a UFA can be converted into the Chrobak Normal Form UFA
$U'$ in \textbf{NLOGSPACE}, though the number of states may go up
polynomially in this conversion  (see Proposition~\ref{ufa-to-cnf-nspace}
below).  
As the bits of the above UFA $U'$ can
be computed as needed, Theorem~\ref{th:logspace} implies the claim.

\begin{propo}
Suppose $U$ is a UFA.
Then, in any accepting run of $U$, there cannot be two distinct simple cycles
(where two simple cycles are considered distinct if they cannot be obtained
from each other by rotation).
\end{propo}
\begin{proof}
If an accepting run contains two distinct simple cycles,
then UFA property is violated by considering repetition of 
these two simple cycles $\ell'$ and $\ell$ times respectively,
where $\ell$ and $\ell'$ are the length of these cycles.
\end{proof}
Let $R_{q,q'}$ denote a path (not necessarily simple)
from state $q$ to $q'$.
Let $R_{q,q'}R_{q',q''}$ denote the
concatenation of the two paths $R_{q,q'}$ and $R_{q',q''}$
to form a path $R_{q,q''}$.

\begin{corollary}\label{ufarun}
Suppose $U$ is a UFA. Then, any accepting run of $U$ 
can be considered as a concatenation
of three paths: $R_{s,q}R_{q,q}^kR_{q,r}$,
where $R_{s,r}=R_{s,q}R_{q,r}$ is a path without
any cycles, and $R_{q,q}^k$ is a simple cycle
from $q$ to $q$ repeated $k$ times.
\end{corollary}
This property is used in the following proposition.
Note that there maybe several such different $R_{s,q}$, $R_{q,r}$
and thus $R_{s,q}R_{q,r}$,
for the same values of $q$ (however, they would be of different
length).

\begin{propo}\label{ufa-to-cnf-nspace}
An $n$ state UFA $U$ can be converted to 
a Chrobak Normal Form UFA $U'$ (having $O(n^2)$ states) 
in \textbf{NLOGSPACE} (note: output tape for $U'$ is not counted
in the space complexity).
\end{propo}
\begin{proof}
The properties of \textbf{NLOGSPACE} used are 
(i) \textbf{NLOGSPACE} is closed under complementation, 
(ii) \textbf{NLOGSPACE} is enough to store constantly many states 
and (iii) \textbf{NLOGSPACE} allows to check whether there is a path from one
state to another in a given (polynomially bounded) number of steps.

Without loss of generality assume that $U$ has only one starting state
(as multiple starting states can be handled separately as the 
set of strings accepted from each starting state is disjoint).
The output Chrobak Normal Form UFA $U'$ will have a stem of size $n$, and
several disjoint cycles following the stem. The $i$-th state of the stem of $U'$
is accepting iff the unary string of length $i$ is accepted by $U$.
Let $s$ be the starting state of $U$. 
For each $j$, with $1 \leq j \leq n$,
put in $U'$ a cycle of size $j$, with $i$-th state in the cycle being accepting
iff the following holds:

\begin{itemize}
\item There exists an accepting state $r$ of $U$, and $h <n$
such that the following conditions are satisfied,
\begin{enumerate}
\item There is an accepting run in $U$ from $s$ to $r$
of length exactly $h$. Note that such an accepting run is unique.
Call this accepting run $R1$.
\item There is an accepting run in $U$ from $s$ to $r$
of length exactly $h+j$. Note that such an accepting run is unique.
Call this accepting run $R2$.
\item In $R1$ no state (including $r$) is repeated.
\item In $R2$, there exists at least one state which is repeated two times,
but there is no state which is repeated three or more times.
\item There is a state $q$ in $U$ which is repeated twice in $R2$ such that
$q$ as well as all the states appearing in $R2$ before the first occurrence of
$q$ and all the states appearing in $R2$ after the second occurrence
of $q$ are in $R1$.
\item $n+i \mod j = h \mod j$.
\end{enumerate}
\end{itemize}

\medskip
\noindent
Note that making the $i$-th state in the cycle of size $j$ accepting in $U'$
as above means that $U'$ accepts all strings of length $\ell \geq n$
such that $\ell \mod j = h \mod j$.

Now, it is claimed that above $U'$ is a UFA and accepts the language
recognised by $U$.

Note that if the conditions above are satisfied for some parameters
$j, i$ along with the corresponding witnesses $r,h$,
then $U$ accepts any unary string with length $\ell \geq n$
such that $\ell \mod j= h \mod j$.
Thus, $\Lang(U') \subseteq \Lang(U)$.

Now it is shown that $\Lang(U) \subseteq
\Lang(U')$.
Consider any string $x$ of length $\ell' \geq n$ which
is accepted by $U$.
Then, by Corollary~\ref{ufarun}
the accepting run of $U$ on $x$ is
of the form $R_{s,q}R_{q,q}^kR_{q,r}$,
where $R_{s,q}R_{q,r}$ is without any cycles, and
$R_{q,q}$ is a simple cycle. Let $h$ be the
length of the run $R_{s,q}R_{q,r}$ and $j$ be the length
of the cycle $R_{q,q}$. Now, $\ell' \mod j = h \mod j$.
Also, by the construction above, $U'$ contains all
strings of length $\ell \geq n$
such that $\ell \mod j = h \mod j$. Thus, $U'$ accepts $x$.

Now it is shown that $U'$ is a UFA.
Note that $U'$ does not have two different accepting runs for
strings of length $<n$.
Suppose by way of contradiction that 
$U'$ is not a UFA.
Then there exists $(i,j) \neq (i',j')$ 
such that in $U'$, $i$-th state of the cycle of size $j$ is accepting
and $i'$-th state of the cycle of size $j'$ is accepting, and
for some string $x$, there are accepting runs
ending in $i$-th state of the cycle of length $j$ and $i'$-state
of the cycle of length $j'$ respectively.
Then the length of string $x$ is $n+i+kj=n+i'+k'j'$ for
some natural numbers $k$ and $k'$.
But then there exist $(r,h)$ and $(r',h')$ such that the conditions 1 to
6 above are satisfied respectively for $(i,j)$ and $(i',j')$.
Note that the triples $(r,h,j)$ and $(r',h',j')$ cannot be same (as otherwise
they would give the same $i$ and $i'$ (by condition 6),
contradicting the fact that $(i,j) \neq (i',j')$).
Note that $n+i+kj \mod j = h \mod j$
and $n+i'+k'j'\mod j'=h' \mod j'$ by condition 6.
But this means, $U$ has at least two different accepting paths for $x$:
in which the cycle in the corresponding witness in $R2$ in condition 2
for parameters $(i,j,r,h)$ and $(i',j',r',h')$ respectively 
are repeated $(n+i+kj-h)/j$ and $(n+i'+k'j'-h')/j'$ times respectively.
\end{proof}

\subsection{Other Operations on UFA except Concatenation}

\noindent
Concatenation operation on UFAs can be complex (see Theorem~\ref{th:concat}
below). In this section, other regular operations (not already handled)
are considered.

\begin{propo}
Let $m$ be the number of primes which are at most $n$, and
let these primes be $p_1, p_2, \ldots, p_m$.
Then, $LCM(1,2,\ldots,n)= \Pi_h\ p_h^{\lfloor n/p_h \rfloor}$.

Furthermore, $\sqrt{n}^m \leq \Pi_h\ p_h^{\lfloor n/p_h \rfloor} \leq n^m$.
Thus, 
$LCM(1,2,\ldots,n)$ and $\Pi_h\ p_h^{\lfloor n/p_h \rfloor}$ are
in $n^{\Theta(n/\log n)}=2^{\Theta(n)}$.
\end{propo}

\begin{proof}
Clearly, the prime factors of $LCM(1,2,\ldots,n)$ are at most $n$.
Furthermore, the largest degree of $p_h$ which divides $LCM(1,2,\ldots,n)$ is
$p_h^{\lfloor n/p_h \rfloor}$. 
Thus, $LCM(1,2,\ldots,n)= \Pi_h\ p_h^{\lfloor n/p_h \rfloor}$.

Furthermore, $\sqrt{n}^m \leq \Pi_h\ p_h^{\lfloor n/p_h \rfloor} \leq n^m$.
Proposition now follows from the prime number theorem.
\end{proof}

\begin{propo}[Holzer and Kutrib \cite{HK02}: (a), (b);
{Jir\'askov\'a} and Okhotin \cite{JO18}: (d);
Okhotin \cite{Ok12}: (g);
Raskin \cite{Ra18}: (i);
Yu, Zhuang and Salomaa \cite{YZS94}: (j)]
\label{prop:regularoperationbounds}
Suppose $U_1, U_2, \ldots, U_k$ are UFAs each having at most $n$ states.
Let $c(n)$ denote $n^{\log n +O(1)}$, the bound as given in
Theorem~\ref{th:complement}.

(a) $\bigcap_{i: 1\leq i \leq k}
\Lang(U_i)$ can be recognised by a UFA having 
number of states bounded by the product
of the number of states in $U_1, U_2,\ldots, U_k$.

(b) There are two UFAs $U_1'$ and $U_2'$ each having
at most $n$ states such that any UFA recognising $\Lang(U_1') \cap
\Lang(U_2)'$ has at least $n^2-n$ states.

(c) 
Let $m$ be the number of primes which are at most $n$, and
let these primes be $p_1, p_2, \ldots, p_m$.
There are UFAs $U_1', U_2',\ldots, U_m'$, with
each having at most $n$ states, such that
any UFA recognising $\bigcap_i \Lang(U_i')$ has
at least $LCM(1,2,\ldots,n)$ (which is in $2^{\Theta(n)}$) states.

(d) Suppose $\Lang(U_1)$ and $\Lang(U_2)$ are disjoint.
Then, $\Lang(U_1) \cup \Lang(U_2)$ can be recognised by a UFA 
having at most $2n$ states.

(e) There are UFAs $U_1'$ and $U_2'$, each having at most
$n$ states and recognising disjoint languages, such that
any UFA recognising their union has at least $2n-4$ states.

(f) $\Lang(U_1) \cup \Lang(U_2)$ can be recognised by a UFA 
having at most $n\cdot c(n)+n$ states.

(g) Let $m$ be the number of primes which are at most $n$, and
let these primes be $p_1, p_2, \ldots, p_m$.
There are UFAs $U_1', U_2',\ldots, U_m'$, with
each having at most $n$ states, such that
any UFA recognising $\bigcup_i \Lang(U_i')$ has
at least $LCM(1,2,\ldots,n)$ (which is in $2^{\Theta(n)}$) states.

(h) Symmetric difference of $\Lang(U_1)$ and $\Lang(U_2)$
can be recognised by a UFA having at most $2n \cdot c(n)$ states.

(i) There are UFAs $U_1'$ and $U_2'$, each having at most
$n$ states, such that any UFA recognising their symmetric difference
needs at least $n^{(\log \log \log n)^{\Omega(1)}}$
states. $U_1'$ can be selected to be
the one-state automaton accepting all words.

(j) For a UFA (of size $n$) for a language $L$ over the
unary alphabet, the language $L^*$ and $L^+$ 
can be recognised by a DFA of size $O(n^2)$.
Furthermore, there exists a DFA (and thus a UFA) $U$ 
having $n$ states, such that any UFA recognising
$\Lang(U)^*$ has at least $\Omega(n^2)$ states.

(k) Consider any number of UFAs $U_1, U_2, \ldots, U_k$,
each having size at most $n$.
There exist two UFAs of size at most $n+LCM(1,2,\ldots,n)$,
which respectively recognise $\bigcap_i \Lang(U_i)$
and $\bigcup_i \Lang(U_i)$.
Note that 
$n+LCM(1,2,\ldots,n)$ is in $2^{\Theta(n)}$.

(l) If one allows regular operations such as Kleene star, Kleene plus
and the Boolean set-theoretic operations (but no concatenation),
then the output of constant-size expressions, with parameters
being given by languages recognised by $n$-state UFAs, can be recognised
by UFAs of quasipolynomial size.

Furthermore, Boolean-valued constant-sized quantifier-free formula with
same type of parameters and comparing such
subexpressions by $=$, $\subseteq$ and $\neq$ can be evaluated
in quasipolynomial time.
\end{propo}

\begin{proof}
Holzer and Kutrib \cite{HK02} showed part (a) by using the standard
product automaton construction which preserves UFA property.
Holzer and Kutrib \cite{HK02} also showed part (b).

Jir\'askov\'a and Okhotin \cite{JO18} showed part (d), as simple union of the
two UFAs recognising the disjoint languages would be a UFA.
Raskin \cite{Ra18} showed part (i), and
Yu, Zhuang and Salomaa \cite{YZS94} showed part (j).

The remaining parts are shown below.

(c) Let $U_1',U_2',\ldots,U_m'$ be DFAs, and thus UFAs, each consisting
of just one cycle of length $p_h^{\lfloor n/p_h \rfloor}$, respectively,
with all states except for the starting state being rejecting.

Then the intersection of languages recognized by the above UFAs is
$H_{okh} =\{1^k: k \mod p_h^{\lfloor n/p_h \rfloor}=0$, for $h=1,2,\ldots, m\}$.
Note that any NFA, and thus UFA, for $H_{okh}$ cannot have a cycle, with
at least one accepting state, of size less than
$\Pi_h\  p_h^{\lfloor n/p_h \rfloor}$.
Thus, the number of states in any NFA/UFA recognizing $H_{okh}$ must 
be at least
$\Pi_h\  p_h^{\lfloor n/p_h \rfloor}=LCM(1,2,\ldots,n)$.

(e) Let $m = \lfloor{n/2}\rfloor$.
Consider a UFA $U_1'$ which consists of only one cycle, which is
of length $2m$ and accepts all strings
with length $0 \mod 2m$. Consider another UFA $U_2'$ which consists of only one
cycle, which is of length $2m-2$
and accepts all strings
with length $1 \mod 2m-2$. 

Now, consider any UFA $U$ accepting $\Lang(U_1') \cup \Lang(U_2')$.
All cycles in $U$ having at least one
accepting state must be of length at least
$2m-2$. This holds, as otherwise $U$
would accept at least one string with length
in the interval $[\ell, \ell+2m-4]$ for every large enough $\ell$. However,
$\Lang(U_1') \cup \Lang(U_2')$ does not contain strings 
in the interval $[k(2m-2)(2m)+2,k(2m-2)(2m)+2m-2]$, for every $k$.

Furthermore, $U$ cannot contain an odd length cycle of size $2m-1$,
as otherwise,
it will accept two even length strings which are $4m-2$ apart: however
$\Lang(U_1') \cup \Lang(U_2')$ contains no such pair of even length strings.
Now, if the UFA $U$ has at least two cycles of different length,
then part (e) follows. If $U$ has cycles of only one length,
then its length must be at least $2m(2m-2)$ due to the smallest eventual
period of 
the characteristic function of $\Lang(U_1') \cup \Lang(U_2')$ being of length
$2m(2m-2)$.

(f) Note that $L \cup H = L \cup (H \cap \overline{L})$.
Part (f) now follows by using parts (a), (d) and Theorem~\ref{th:complement}.

(g) Let $U_1',U_2',\ldots,U_m'$ be DFAs, and thus UFAs, each consisting
of just one cycle of length $p_h^{\lfloor n/p_h \rfloor}$, respectively,
with all states except for the starting state being accepting.

Then the union of the languages recognized by the above UFAs is
$L_{okh} =\{1^k: k \mod p_h^{\lfloor n/p_h \rfloor} \neq 0$, for some
$h \in \{1,2,\ldots, m\}\}$.
By Okhotin \cite[Lemma 5]{Ok12} any UFA recognizing it must have at least
$\Pi_h\  p_h^{\lfloor n/p_h \rfloor}=LCM(1,2,\ldots,n)$
states.

(h) Note that symmetric difference of $L$ and $H$ is $(L \cap \overline{H})
\cup (H \cap \overline{L})$.
Part (h) now follows by using parts (a), (d) and Theorem~\ref{th:complement}.

(k) Consider arbitrary number of UFAs $U_1, U_2, \ldots, U_k$
each having at most $n$
states. Translate each of these UFAs into Chrobak normal form, without
size increase. Then, extend the stem of these UFAs such that it has exactly
$n$ states, to make the stem of all the UFAs of the same length.
The cycles after these $n$ states are all of length at most $n$.
Call these UFAs $U_1', U_2', \ldots,U_k'$.
These UFAs can then be translated into DFAs $A_1, A_2, \ldots, A_k$,
which recognise the same languages as $U_1, U_2, \ldots, U_k$ respectively.
Each of the $A_i$'s have a stem of length $n$, followed by a cycle whose length is 
$LCM(1,2,\ldots,n)=\Pi_h\  p_h^{\lfloor n/p_h \rfloor}$.
The stem of $A_i$ is the same as the stem of $U_i'$. The $j$-th state in
the cycle of $A_i$ is accepting iff there exists a cycle, say of length $\ell$,
in $U_i'$ such that $j \mod \ell$-th state in it is accepting.

Now, let $A$ be a DFA which has a stem of size $n$, and a cycle
of size $LCM(1,2,\ldots,n)$.
The $j$-th state in
the stem (cycle) is accepting iff the $j$-th state
in the stem of each of the $A_i$'s (cycle of each of the $A_i$'s) is accepting.
Then, $\Lang(A)=\bigcap_i Lang(A_i)$.

If instead, one has that the $j$-th state in
the stem (cycle) is accepting iff the $j$-th state
in the stem of one of the $A_i$'s (cycle of one of the $A_i$'s) is accepting.
Then, $\Lang(A)=\bigcup_i Lang(A_i)$.

The size of DFA $A$ above is bounded by $n+LCM(1,2,\ldots,n)$.

(l) This follows from the above parts and Theorem~\ref{th:complement},
as composition of constant number of quasipolynomial bounds
gives a quasipolynomial bound.
\end{proof}

\noindent
The bounds for arbitrary unions and intersections of $n$-state DFAs
are the same as those for UFAs. The lower bounds combine
$\Theta(n/\log n)$ DFAs and the upper bounds produce a UFA
which is at the same time a DFA. For the NFAs, the bounds
for the intersection only change by an additive term $O(n^2)$.
This holds, as the common stem of the Chrobak Normal Forms is of that 
length, while the loops which can occur are the same as in
the UFA case. For the union, the lower bounds do not carry
over. Instead one can transform all NFAs into NFAs with a
stem of the same length -- it has, independently of the number
of input NFAs, quadratic length -- followed by cycles of
lengths up to $n$. Let $N_1,\ldots,N_k$ be the input NFAs
in this joint Chrobak Normal Form. Then for the output NFA
$N$, a node of the stem of $N$ is accepting if and only if, in one
of the input $N_h$, the node at the same position is accepting.
A node in a cycle of length $\ell$, where $\ell = 1,2,\ldots,n$,
is accepting if and only if, for some $N_h$ having a cycle of
length $\ell$, the node at the same position in the cycle of
length $\ell$ of $N_h$
is accepting. Thus the overall size of this resulting NFA
$N$ is $O(n^2)$.

The following proposition considers space complexity of the 
various operations (except concatenation) on UFAs.

\begin{propo} \label{prop:quasipolynomial}
Given UFAs $U_1$ and $U_2$ with at most $n$ states as input,
a \textbf{POLYLOGSPACE} algorithm can output a UFA which
recognises the intersection and union of the languages
$\Lang(U_1)$ and $\Lang(U_2)$, as well as the Kleene Star
or complement of the language $\Lang(U_1)$.
\end{propo}
\begin{proof}
Note that, as usual in computational complexity,
the space count only refers to the work tape
and not to input and output tapes.

Intersection: UFA for $\Lang(U_1) \cap \Lang(U_2)$
can be constructed by using the product automaton
of $U_1$ and $U_2$. This product automaton
has states of the form $(q_1,q_2)$ such that $q_i$ is a state of $U_i$.
The starting states (accepting states) of the automaton are $(q_1,q_2)$ such that
$q_i$ is a starting state (accepting state) of $U_i$.
The transition function is $\delta((q_1,q_2),a)=(\delta_1(q_1,a), \delta_2(q_2,a))$,
where $\delta_i$ is the transition function of $U_i$.
The above is clearly doable in \textbf{LOGSPACE}.

Union: $\Lang(U_1) \cup \Lang(U_2)=
	\overline{\overline{\Lang(U_1)} \cap \overline{\Lang(U_2)}}$. 
Thus, a required UFA can be found using
the intersection algorithm given above
and the complementation algorithm given below.

Kleene Star: Yu, Zhuang and Salomaa \cite{YZS94} showed that the
Kleene star of a unary language given by an $n$-state NFA can be recognised
by a DFA of size $(n-1)^2+1$. Note that this DFA consists of a stem, followed
by a cycle.
Thus, one can search using an \textbf{NLOGSPACE}
algorithm for an $m$, with $1\leq m \leq n^2$, such that, 
for $k=0,1,\ldots,(n+1)^4$, every word of length
$n^2+k$ is accepted iff the word of length
$n^2+m+k$ is accepted. This length $m$ is then the period which starts
latest at length $n^2$.
Thus, the algorithm can output a DFA of size $O(n^2)$
which consists of a stem of length $n^2$ followed by a cycle of length $m$,
where \textbf{NLOGSPACE} is enough to detect if each of the states involved
is accepting.

Complementation: For complementation, the handling of the stem is standard
(just convert rejecting states to accepting and accepting states to rejecting).
So, below assume that the UFA $U_1$ consists only of cycles. 
The basic idea is to use the algorithm in Theorem~\ref{th:complement}.
This algorithm has mainly
two running variables for the recursive descent: $d$ and $k$.
Both take at most the value $n^{(\log n)/2+c}$ for some constant $c$
and therefore can be written with $O(\log^2 n)$ bits. Furthermore,
during recursion, one has to archive the old values before branching.
Thus, the algorithm archives $(d_{h'},k_{h'},e_{h'},s_{h'})$ from the
algorithm in Theorem~\ref{th:complement} for each level $h'$ and this
can be done with $O(\log^3 n)$ space.
Furthermore, note that the transformation into the Chrobak Normal Form also
takes just $O(\log^2 n)$ space as by Savitch's Theorem
$O(\log n)$ nondeterministic space is contained
in $O(\log^2 n)$ deterministic space.
Thus, the algorithm from Theorem~\ref{th:complement} 
is a \textbf{POLYLOGSPACE} algorithm.

The above space bound can be improved somewhat by noting the following:
instead of archiving the full tuples $(d_{h'},k_{h'},e_{h'},s_{h'})$,
one could archive
the index information $e_{h'},s_{h'}$ when going from level $h'$ to $h'+1$,
along with the quotient $d_{h'+1}/d_{h'}$.
When returning from level $h$ to level $h-1$,
one computes $d_{h-1} = d_h / (d_h/d_{h-1})$
where $d_h/d_{h-1}$ was archived for level $h-1$ and furthermore, one computes
$k_{h-1} = k_h - s_{h-1} \cdot d_{h-1}$.
With these modifications, the algorithm runs in $O(\log^2 n)$ space.
\end{proof}

\subsection{Concatenation}

\noindent
It is well-known that the concatenation of two $n$-state NFAs can be
realized by a $2n$ states NFA.
Furthermore, Pighizzini \cite{Pi00,PS02} showed that the concatenation
of two unary DFAs of size $n$ can be realised by a DFA of size $O(n^2)$.
Pighizzini's result allows for an implementation of the
following concatenation algorithm 
for UFAs:
Convert the two UFAs into DFAs \cite{Ok12} and then apply the algorithm 
for the concatenation of DFAs. This gives the upper bound of
$2^{O((n \log^2 n)^{1/3})}$ on the size of a UFA realising the 
concatenation of two UFAs of size $n$.
The following theorem gives
a matching lower bound.

\begin{theorem} \label{th:concat}
There is an exponential-type blow-up for UFA sizes when
recognising the concatenation of unary languages. The concatenation
of two languages given by $n$-state UFAs requires, in the
worst case, a UFA with $2^{\Omega((n \log^2 n)^{1/3})}$ states.
\end{theorem}

\begin{proof}
Let $m$ be a numeric parameter and consider the first $k=m-2$ primes,
$p_0,\ldots,p_{m-3}$ which are all at least $m$.
Note that each $p_i \leq c \cdot m\log m$,
for some constant $c$, by the prime number theorem.
Now the UFA $U$ to be constructed contains
$k$ cycles $C_{\ell}$
of length $p_\ell \cdot m$, for $\ell = 0,1,\ldots,k-1$.
The cycle $C_\ell$ has, 
for $h=0,1,\ldots,p_\ell-2$:
$(\ell+1+h \cdot m))$-th states
as accepting. The $0$-th state in the cycles are the starting state of the cycle.
There is one further cycle $C'$ of length
$m$, which has $0$-th state as starting start and $1$-th state as accepting state.
Let $L$ denote the language recognised by this UFA. 

The lengths of $m-1$ consecutive unary strings not being accepted
by the above UFA are exactly at lengths
$r \cdot m + 2, \ldots, r \cdot m+m$, where $r$ is
$p_\ell -1$ modulo $p_\ell$ for each $\ell = 0,1,\ldots,k-1$. Furthermore, this does
not happen at any other lengths. To see this note that:
(i) Cycle $C'$ accepts all strings of length $1$
modulo $m$. Thus, $m-1$ consecutive
unary strings not being accepted by the above UFA can happen
only if these consecutive strings are of length
$r \cdot m +2, \ldots, r \cdot m+m$ for some $r$.
(ii) For $\ell < k$, cycle $C_{\ell}$,
does not accept the string of length $r \cdot m+2+\ell$ iff
$r \mod p_{\ell}=p_{\ell}-1$.
It follows from the above two statements that $m-1$ consecutive unary
strings not being accepted by the above UFA happens iff these strings
are of length
$r \cdot m+2, \ldots, r \cdot m+m$, for some $r$ such that, for
all $\ell<k$, $r \mod p_{\ell}=p_{\ell}-1$.
Note also that $L$ contains length $1$ word.

Let $H$ be the finite language which contains the words of length
$0,1,\ldots,m-2$ and no other words.
Now $L \cdot H$ contains
all words except those whose length is of the form $r \cdot m $, where
$r \mod p_{\ell}= 0$, for each $\ell=0,1,\ldots, k$.

Note that, by Okhotin \cite[Lemma 5]{Ok12}, any UFA recognising the
language $L \cdot H$ has at least  $\Pi_i \ p_i \geq m^{m-2}
=2^{\Omega(m \log m)} $ states.

Let $n$ be the number of states in the above UFA.
Then, $n = \Theta((m-2) \cdot m \cdot m \log m+m)= \Theta(m^3 \log m)$,
and thus, $m = \Theta((n/\log n)^{1/3})$.
It follows that a UFA for $L \cdot H$ needs at least
$\Theta((n\log^2 n)^{1/3})$ states.
\end{proof}

\noindent
Next one considers the evaluation of constant-sized formulas
using UFAs as input.
Subsequent results provide a lower bound of $2^{\Omega'((n\log n)^{1/3})}$ for
the time complexity to evaluate formulas (including concatenation)
over languages recognised by UFAs.
This bound almost matches the upper bound 
obtained by transforming the UFAs into DFAs ($2^{\Theta((n \log ^2 n)^{1/3}})$
given by \cite{Ok12}), and
then using the polynomial
bounds on the size increase of DFAs over the unary alphabet for regular operations.

In the following, $H_1,H_2,K,L$ are sets of
words given by $n$-state UFAs and $K$ is a finite language.

\begin{theorem} \label{thm:ethformula}
Assuming the Exponential Time Hypothesis,
it needs time $2^{\Omega'((n\log n)^{1/3})}$ to
evaluate the truth of the formula
$$
	(H_1 \cap H_2) \cdot K = L
$$
where $H_1,H_2,K$ are given by UFAs with at most $n$-states and 
$L$ is the set of all words.
\end{theorem}

\begin{proof}
Consider a $3$SAT formula with clauses $c'_1,c'_2,\ldots,c'_m$, where
each variable appears in at most $3$ clauses.
Divide the clauses into $r=\lceil m/\log m \rceil$
disjoint groups of $\log m$ clauses each
(where the rounded down value of $\log m$ is used).
For $i<r$, group $G_i$ has the clauses $c'_{i \cdot (\log m) +1},
\ldots, c'_{(i+1) \cdot (\log m)}$
(the last group $G_{r-1}$ may have fewer clauses due
to $m$ not being divisible by $\log m$).

If a variable, say $x$, appears in different groups of clauses, then one
can rename $x$ in different groups to $x', x'', \ldots$ and
add equality clauses $(x'=x''), \ldots$. 
Thus, by adding some additional equality clauses,
it can be assumed that no variable appears in two different
group of clauses.
Note that there can be at most $O(m)$ equality clauses.

So, now the $3$SAT formula has the clauses $c_1, c_2, \ldots, c_m$ (which
have at most 3 literals each)
and the equality clauses $c_{m+1}, c_{m+2}, \ldots, c_{m'}$.
The clauses $c_1, c_2, \ldots, c_m$ are divided into groups
$G_0,G_1, \ldots, G_{r-1}$ containing
at most $\log m$ clauses each,
and any variable appears in clauses from at most one group, and
perhaps in the equality clauses.

As each group contains at most $\log m$ clauses, and thus at most $3 \log m$
variables, there are at most $8m$ possible truth assignments
to variables appearing in clauses of any group.
Below, $k$-th truth assignment (starting from $k=0$) to variables assigned
to $p_i$ assumes some ordering among the truth assignments where
if the number of truth assignments is at most $k$, then
$k$-th truth assignment is assumed to be the $0$-th truth assignment
(the latter part is just for ease of writing the proof).

Consider $r$ distinct primes $p_0, p_1, \ldots p_{r-1}$,
each greater than $8m$ but below some constant $c$ times $m$. 
Note that, for large enough constant $c$, there exist
such distinct primes by the prime number theorem.

Assign variables / clauses appearing in group $G_i$ to prime $p_i$.
Now UFA for $H_1$ consists of the $r$ cycles $C_0,C_1,\ldots,C_{r-1}$ 
and UFA for $H_2$ consists of the $r$ cycles $C'_0,C'_1,\ldots,C'_{r-1}$.
The cycles $C_i$ and $C'_i$ are of length $2(m'+1) \cdot p_i$.
For $k<p_i$, and $j<m'$,
the $k \cdot (2m'+2)+2j$-th and $k \cdot (2m'+2) + 2j+1$-th states in
cycle $C_i$ and $C'_i$ are accepting or non-accepting based on
the table given in Figure~\ref{theorem13}.
Intuitively, for $c_j$ assigned to $p_i$, $C_i$ and $C_i'$ will test for
satisfaction (4th row in the table below).
If $c_j=(x=y)$, and $x$ is assigned to $p_i$ and $y$ to $p_{i'}$, 
then 5th and 6th rows in the table below for $C_i$ and $C_{i'}'$
respectively
will check for consistency in the assignment to the variables $x$ and $y$.
In the table of acceptance and rejection (Figure~\ref{theorem13})
for the entries of the cycles $C_i$ and
$C_i'$, the value $i$ can be considered as constant and the
value $j$ is the running variable going over all clauses.

\begin{figure}[t]
\caption{Table of Finite Automaton.}
\label{theorem13}
\begin{center}
\begin{tabular}{|c|c|c|}
\hline
 & $k \cdot (2m'+2)+2j$-th & $k \cdot (2m'+2)+2j+1$-th \\
 & state in $C_i$ / $C_i'$ & state in $C_i$ / $C_i'$  \\ \hline
$j=0$, $i=0$ &   accepting &   accepting \\  \hline
$j=0$, $i \neq 0$  &   not-accepting  &   not-accepting \\  \hline
$0<j\leq m$ and & accepting iff & accepting iff \\
all variables of $c_j$
& $c_j$ not satisfied by the $k$-th & $c_j$ not satisfied by the $k$-th \\
belong to $G_i$ & truth assignment to variables & truth assignment to variables \\
        & assigned to $p_i$ & assigned to $p_i$ \\ \hline
This row is applicable & accepting iff truth value & accepting iff truth value \\
only for $C_i$	where & assigned to $x$ is true & assigned to $x$ is false \\
	$m <j \leq m'$ and
	& in the $k$-th truth assignment &  in the $k$-th truth assignment \\
	$c_j$ is $(x=y)$ and
  & to the variables assigned to $p_i$ & to the variables assigned to $p_i$         \\
	$x$ is assigned to $p_i$  & & \\ \hline
This row is applicable & accepting iff truth value & accepting iff truth value \\
only for $C_i'$ where & assigned to $y$ is false & assigned to $y$ is true \\
$m <j \leq m'$ and & in the $k$-th truth assignment &  in the $k$-th truth assignment \\
	$c_j$ is $(x=y)$ and
  & to the variables assigned to $p_i$ & to the variables assigned to $p_i$         \\
	$y$ is assigned to $p_i$ & & \\ \hline
All other cases & not accepting & not accepting \\ \hline
\end{tabular}
\end{center}
\end{figure}

Note that the above automatons are unambiguous:
(i) for $j=0$, only cycle $C_0,C_0'$ could accept,
(ii) for $1 \leq j \leq m$, only $C_i,C_i'$ such that $c_j$ is assigned 
to $p_i$ can accept, and
(iii) for $m<j \leq m'$, if $c_j$ is $(x=y)$ with
$x$ and $y$ assigned to $p_i$ and $p_{i'}$ respectively,
only $C_i$ and $C'_{i'}$ respectively can accept.

Note that all strings with length being $0$ or $1$ modulo $(2m'+2)$
are in $H_1 \cap H_2$ (due to the 2nd row in the definition
for $C_i$ and $C_i'$, for $j=0$ and $i=0$).

Now, if the $3$SAT formula is satisfiable, then consider some satisfying
truth assignment to the variables, say it is the $k_i$-th truth assignment
to variables assigned to $p_i$.
Then, for any $s$ such that for $i<r$, $k_i = s \mod p_i$, 
$H_1 \cap H_2$ will not contain strings of length $s(2m'+2) +2j$ and $s(2m'+2)+2j+1$ for
$1 \leq j \leq m'$:
(a)
if $1 \leq j \leq m$, clause $c_j$ is satisfied and thus these strings are
not accepted by $H_1$ and $H_2$
(see 4th row in the definition of $C_i$ and $C_i'$, for $c_j$ being assigned to $p_i$), and
(b) if $m < j \leq m'$ and $c_j=(x=y)$, where $x$ is assigned to $p_i$ and $y$ to $p_{i'}$,
then as the variable assignments are consistent again these strings are accepted by
only one of $H_1$ and $H_2$ (see 5th and 6th rows in the definition of 
$C_i$ and $C_{i'}'$ respectively).
Thus, $H_1 \cap H_2$ misses $2m'$ consecutive strings.

On the other hand, if $H_1 \cap H_2$ misses $2m'$ consecutive strings, then it must be
strings of length $s(2m'+2)+j$ for $2 \leq j \leq 2m'+1$, for some value of $s$ (as all strings with lengths
being $0$ or $1$ modulo $2m'+2$ are in $H_1 \cap H_2$). Then let
$k_i =s \mod p_i$. Then, 
for the $k_i$-th assignment of truth values to the variables assigned to $p_i$, it must be the case that
all the clauses $c_j$ assigned to $p_i$ are satisfied
(otherwise by the 4th row in the definition of $C_i$, $C_i'$, it can be seen that
$H_1$ and $H_2$ will contain $s(2m'+2)+2j$ and $s(2m'+2)+2j+1$).
Furthermore, the equality clauses are satisfied. To see this, suppose
some equality clause $c_j=
(x=y)$ is not satisfied, where $x$ is assigned to $p_i$ and $y$ is assigned to $p_{i'}$.
Then by the 5th and 6th rows in the definitions of $C_i$ and $C_{i'}$ respectively, 
one of $s(2m'+2)+2j$ and $s(2m'+2)+2j+1$ length strings is in both $H_1 \cap H_2$
(depending on
the truth assignment to $x$ and $y$ in the $k_i$-th and $k_{i'}$-th truth
assignments to the variables assigned to $p_i$ and $p_{i'}$ respectively).

Thus, $H_1 \cap H_2$ misses out on $2m'$ consecutive strings iff the given
$3$SAT formula is satisfiable.
Taking $K$ to be set of strings of length $0,1,2,\ldots, 2m'-1$, gives us that
$(H_1 \cap H_2) \cdot K$ is universal iff $3$SAT formula is not satisfiable.

Let $n$ denote the size of the UFAs for $H_1,H_2$.
Note that the size of $K$ is bounded by $2m' \leq n$.
Now, $n \leq (2m'+2) \cdot (\sum_i p_i)=\Theta(m^3/\log m)$.
Thus, $n\log n=\Theta(m^3)$ or $m=\Theta((n\log n)^{1/3})$.
The theorem now holds assuming ETH.
\end{proof}

\noindent
Note that Okhotin \cite{Ok12} provides an upper bound 
for evaluating formulas over UFAs by first converting
UFAs into DFAs and then carrying out the operations with DFAs. These
operations run in time polynomial in the size of the DFAs constructed.

\begin{theorem} \label{th:compbound}
Under the assumption of the Exponential Time Hypothesis:

	It takes time at least $2^{\Omega'(n^{1/4})}$ 
to decide if the concatenation of the languages recognised
by two given UFAs of at most $n$ states is universal.
\end{theorem}

\begin{proof}
This proof uses the construction used in Theorem~\ref{thm:ethformula}.
Consider a $3$SAT formula with clauses $c'_1,c'_2,\ldots,c'_m$, where
each variable appears in at most $3$ clauses.
As in the proof of Theorem~\ref{thm:ethformula} construct updated
SAT formula which contains clauses
$c_1, c_2, \ldots, c_m$ (which have at most 3 literals each) and the equality clauses
$c_{m+1}, c_{m+2}, \ldots, c_{m'}$.
Then construct $H_1$, $H_2$ and $K$ as in the proof of Theorem~\ref{thm:ethformula}. 

The cycles of $H_1$ and $H_2$
in the proof of Theorem~\ref{thm:ethformula} have
length $2(m'+1) \cdot p$ for some number $p$ of size $O(m)$
which is either a prime or the constant $1$. Furthermore, for each
number $\ell=0,1,\ldots,2m'+1$, there is at most one cycle $A_\ell$ in 
the UFA for $H_1$
and at most one cycle $B_\ell$ in the UFA for $H_2$
which accepts a string of length $\ell$ modulo $2(m'+1)$ (if any).

Let lcm$(a,b)$ denotes the least common multiple of $a$ and $b$.
Now construct a new intersection automaton of $H_1$ and $H_2$ as follows.
It consists of $2(m'+1)$ cycles $E_\ell$, $\ell \leq 2m'+1$. $E_\ell$
has the length lcm$(|A_\ell|,|B_\ell|)$. The $t$-th state of $E_\ell$
is accepting iff $t$ has the remainder $\ell$ when divided by $2(m'+1)$
and both cycles $A_\ell$ and $B_\ell$, after 
$t$ steps from the start state of those cycles, are in an accepting state
in the corresponding automata $H_1$ and $H_2$.
Thus, the constructed intersection automaton consists of all cycles
$E_\ell$ and accepts a word iff both $H_1$ and $H_2$ accept this word.
Furthermore, the automaton, for each $\ell=0,1,\ldots,2m'+1$, for any $h$,
accepts a string of length $\ell+2(m'+1)h$,
only in the cycle $E_\ell$, if at all. Thus, the constructed
automaton is a UFA.

As the size of each cycle $E_\ell$ is $O(m^3)$ and as there are
$2(m'+1)$ cycles and $m' \in \Theta(m)$, the overall size of the above
constructed automaton (call it $H$) is $O(m^4)$. 
Taking $K$ as in the proof of 
Theorem~\ref{thm:ethformula}, it follows that $H \cdot K=(H_1 \cap H_2)
\cdot K$ is universal iff $3$SAT formula is not satisfiable.
Let $n$ denote the size of $H$. Then, $m=\Theta(n^{1/4})$.
Note that $n$ also bounds the size of $K$.
Thus, the universality problem for the concatenation of
two $n$-state UFAs is in $2^{\Omega'(n^{1/4})}$ under the assumption
of the Exponential Time Hypothesis.
\end{proof}

\noindent
As a UFA is also an NFA, a
$2n$ state NFA for the concatenation of the languages recognised
by two $n$-state NFAs can be constructed in polynomial time.
Theorem~\ref{thm:algo} proves that the universality
of an $n$ state NFA can be checked in time $2^{O((n \log n)^{1/3})}$.
As $O((2n \log 2n)^{1/3}) = O((n \log n)^{1/3})$,
one has the following corollary to Theorem~\ref{thm:algo}.
Note that this bound is slightly better than the method of
converting the UFA to DFA and doing the concatenation of
DFAs as done by Okhotin \cite{Ok12}.

\begin{corollary}
It can be decided in time $2^{O((n \log n)^{1/3})}$ whether the
concatenation of the languages recognised by two given UFAs is universal.
\end{corollary}

\section{Membership in Regular Languages of Infinite Words}

\noindent
For unary alphabet, the $i$-th word is the word of length $i$.
Let $L(i)=1$ if the $i$-th word is in $L$, else $L(i)=0$.
Thus, the characteristic function $L(0)L(1)\ldots$ of the language $L$ 
can be viewed as an $\omega$-word generated by the language $L$
(or generated by the NFA/UFA accepting $L$).

An $\omega$-language is a set of $\omega$-words and it is called $\omega$-regular
iff a nondeterministic B\"uchi automaton recognises the language.

B\"uchi \cite{Bu60} showed in particular that an $\omega$-language
is $\omega$-regular iff there exist finitely many regular languages
$A_1,A_2,\ldots,A_n,B_1,B_2,\ldots,B_n$ such that
$B_1,B_2,\ldots,B_n$ do not contain the empty word and
the given $\omega$-language equals $\bigcup_k A_k \cdot B_k^\infty$
where $B_k^\infty$ is the set of infinite concatenations of members in
$B_k$. Note that all elements of an $\omega$-language are $\omega$-words.

Now the question investigated is the following: Given a fixed
$\omega$-regular language, what is the complexity to check whether
the $\omega$-word generated by a unary $n$-state NFA is in this
given $\omega$-regular language. The lower bound of Fernau and Krebs \cite{FK17}
(see Proposition~\ref{propone}) shows that, 
assuming the Exponential Time Hypothesis, 
by choosing the fixed $\omega$-regular language as $1^*$,
this task requires 
at least $2^{\Omega'(n^{1/3})}$ deterministic time,
where $n$ is the number of states of the input NFA.
The following theorem gives a better lower bound.

For the following proof, it is useful to consider a simple cycle in an
NFA/UFA of length $n$ as coding
a length $n$ word $w \in \{0,1\}^*$, where the $i$-th character in $w$ is
$1$ iff the $i$-th state in the cycle is accepting.
Thus, if an NFA consists of just one simple cycle (and no stem), then the
$\omega$-word generated by the NFA is $w^{\infty}$, where
$w$ is the word coded by the cycle in the NFA.
Correspondingly, the $\omega$-word generated by an NFA with several disjoint
simple cycles
(and no stem), can be viewed as an $\omega$-word formed by taking
bitwise or of the $\omega$-words generated by the individual cycles.

\begin{theorem} \label{th:nfaomega}
(a) Assuming the Exponential Time Hypothesis, checking whether an $n$-state
NFA defines an $\omega$-word in a fixed $\omega$-regular
language $\cal L$
takes at least time $2^{\Omega'((n \log\log n / \log n)^{1/2})}$.

(b) Checking whether an $n$-state
NFA defines an $\omega$-word in a fixed $\omega$-regular $\cal L$
can be done in time $2^{O((n \log n)^{1/2})}$.
\end{theorem}

\begin{proof}
Impagliazzo, Paturi and Zane \cite{IPZ01} showed that whenever the
Exponential Time Hypothesis holds, then this is witnessed by a sequence
of $3$SAT formulas which has linearly many clauses when measured by the
number of variables. Such clauses will be coded up as follows in an NFA.

Code an $m$ variable, $k$ clause $3$SAT with $k\in O(m)$
using an NFA as follows.
Suppose $x_1,\ldots,x_m$ are the variables used in the $3$SAT formula,
and $c_1,\ldots, c_k$ are the $k$ clauses.

Let $r'=\ceil{\log \log m}$, $r=2^{r'}$.
Without loss of generality assume $m$ is divisible by $r'$:
otherwise one can increase $m$ (upto doubling) to make this hold
and add one literal clauses to the $3$SAT formula
using the new variables, without changing the complexity stated in the theorem.

The NFA has $t=\frac{m}{r'}$ disjoint cycles of different prime lengths
$p_1,p_2,\ldots,p_t$, and
each of these prime numbers are at least $4rk+14r+1$ and in
$\Theta(m \log m)$. Note that
by the prime number theorem, as $k \in O(m)$, there are such different
primes, where the constant in the $\Theta$ also depends on the constant
of the linear bound of $k$ in $m$.
So the overall size of the NFA is $\Theta(m^2 \log m/ \log \log m)$. 
Intuitively, a cycle of length $p_i$ is used to handle the
variables $x_{(i-1) \cdot r'+1}$
$x_{(i-1) \cdot r'+2}$ $\ldots$ $x_{(i-1) \cdot r'+r'}$.

For each prime length $p_i$, where $i=1,2,\ldots,t$ as above, 
the cycle of length $p_i$ codes
$(1100000(10a_{i,\ell,h}0)_{h=1}^{k}1000011)_{\ell=1}^{r}1^{p_i-r(4k+14)}$
where $1$ denotes accepting state in the cycle and $0$ denotes rejecting
state in the cycle.
Here $a_{i,\ell,h}$ is $1$ if the 
truth assignment to $x_{(i-1) \cdot r'+1}$
$x_{(i-1) \cdot r'+2}$ $\ldots$ $x_{(i-1) \cdot r'+r'}$
being the $\ell$-th binary string (starting with $\ell=1$)
in $\{0,1\}^{r'}$
makes the $h$-th clause $c_{h}$ true, and $0$ otherwise.
Each portion $(1100000(10a_{i,\ell,h}0)_{h=1}^{k}1000011)$ is called
a block. Each block codes a possible truth assignment to the variables
encoded by prime $p_i$: block corresponding to a value $\ell$ corresponds
to the $\ell$-th truth assignment to the variables 
$x_{(i-1) \cdot r'+1}$ $x_{(i-1) \cdot r'+2}$ $\ldots$ $x_{(i-1) \cdot r'+r'}$.
The part $10a_{i,\ell,h}0$ corresponds to checking
if the $h$-th clause is satisfied by the $\ell$-th truth assignment
to the variables assigned to $p_i$.
Note that $r = 2^{r'}$ is the number of possible truth-assignments
to these $r'$ variables.
Note that five consecutive zeros only occur at the beginning of a block.
Furtheremore, $100001$ occurs only at the end of a block.

Each cycle has a different prime length. Thus,
by the Chinese Remainder Theorem, for each possible truth assignment 
to the variables, there is a number $s$ such that
$s \mod p_i$ is the starting position of the block where the
corresponding variable values are used for evaluating which clauses
are satisfied.
Therefore, if a truth assignment leads to the $3$SAT formula
being true, then for the language $L$ recognised by the NFA,
$L(s)L(s+1)\ldots L(s+4k+13)$ would be $1100000(1010)^k1000011$
which is in $1100000(1010)^+1000011$. 
On the other hand, if 
$1100000(1010)^+1000011$ is a substring of 
$L(0)L(1)\ldots$, then let $L(s)$ be the starting point for
the substring $1100000(1010)^+1000011$ in $L(0)L(1)\ldots$.
Then, as five consecutive $0$s in $11000001$ can 
happen only at the start of a block of any cycle of
the NFA, it must be the case that all the cycles have
a block starting at $s$.
Thus, $(1010)^+$ must be of the form $(1010)^k$. Hence,
for some $\ell_i$ corresponding to each $i$,
for each value of $h$ in $\{1,2,\ldots, k\}$, for some $i$,
$a_{i,\ell_i,h}$ is $1$.
Thus, the $3$SAT formula is satisfiable.

This proves the reduction of a $3$SAT formula to the question
whether the $\omega$-word generated
by the corresponding NFA is in the $\omega$-regular language
$$\{0,1\}^*1100000(1010)^+1000011\{0,1\}^{\infty}.$$
Now $n \in \Theta(m^2 \log m / \log \log m)$ and
therefore
$\Theta(\log n) = \Theta(\log m)$. Furthermore, it follows that
$m \in \Omega((n \log\log n/\log n)^{1/2})$ and,
by the Exponential Time Hypothesis, determining the membership
of the $\omega$-word defined by an unary $n$-state NFA requires at
least time $c^{(n \log\log n/ \log n)^{1/2}}$ for some constant $c$.

(b) For the upper bound, first convert the NFA
to a DFA. Then, compute
words $v,w$ such that the $\omega$-word generated 
by the DFA
equals $v w^\omega$. Note that such $v, w$
can be easily computed for DFA over unary
alphabet, as the DFA would consist of
a stem, and then possibly one cycle.
For checking whether
there is a $k$ such that
$v w^\omega \in A_k \cdot B_k^\infty$,
consider a deterministic Muller automaton for
the given $\omega$-regular language.
Now, one first feeds the Muller automaton $v$
and records the state after processing $v$.
Then one feeds the Muller automaton each time
with $w$ and records the entry and exit states and
which states were visited during the processing of
$w$. This is repeated until there is a repetition of
the entry state, which happens after
processing finitely many copies of $w$.
The states visited in this loop 
after processing of finitely many $w$,
are the infinitely often visited states.
This allows to
test the membership of $v w^\omega$
in $A_k \cdot B_k^\infty$.
The time complexity of this algorithm is in
$2^{\Theta((n \log n)^{1/2})}$ --- 
as the NFA to DFA conversion can be done
in time $2^{\Theta((n \log n)^{1/2})}$,
and thereafter the processing
in Muller Automata is polynomial in this bound.
\end{proof}

\begin{remark}
In the above proof,
instead of the cycle of length $p_i$ coding,
$$(1100000(10a_{i,\ell,h}0)_{h=1}^{k}1000011)_{\ell=1}^{r}1^{p_i-r(4k+14)}$$
as in Theorem~\ref{th:nfaomega},
extend the cycle to code
$$1100011000110001100011 
(1100000(10a_{i,\ell,h}0)_{h=1}^{k}1000011)_{\ell=1}^{r}1^{p_i-r(4k+14)}.$$
This makes the cycle length bigger by 22, and thus the primes correspondingly
need to be at least $4rk+14r+23$.
Note that the 22 bit `marker' $1100011000110001100011$ 
occurs as a sub-word at location $s$ in the $\omega$-word of the NFA
iff all cycles are in the initial position at location $s$ (that is all
cycle lengths divide $s$).
Otherwise, for the cycle $C$ not in the initial
position at location $s$,
some overlay with the marker in $C$, or
with a consecutive run of three $1s$ in
$C$ or the subword $10a010b01$,
(with $(a,b \in \{0,1\})$) in $C$
would provide an additional $1$ to go into one of
these 22 positions following location $s$, contradicting
the occurrence of the marker at location $s$.

Further note that,
by the Chinese Remainder Theorem,
between two consecutive occurrences of the marker,
for each possible combination of $\ell_i$, $1 \leq i \leq r$, there exists a
unique position in the $\omega$-word of the NFA where the match of the portion 
$(1100000(10a_{i,\ell_i,h}0)_{h=1}^{k}1000011)$ for cycle of length $p_i$
happens for all $i$ with $1\leq i \leq t$.

Now fix some constant $q$.
Now an $\omega$-automaton, in the input $\omega$-word,
can recognise two consecutive occurrences of the marker
$1100011000110001100011$.
In between these two occurrences, the automaton can
count, modulo $q$, the number of blocks	
which are of the form $1100000\{1010,1000\}^+1000011$ and have the subword
$10001$. This gives the number of non-solutions (modulo $q$) for 
the $3$SAT instance when the corresponding $\omega$-word for the NFA given
above is used as input.

Therefore, the membership test in the $\omega$-regular language corresponding
to the above $\omega$-automaton,
can check whether the correctly coded 
$\omega$-words representing an $m$-variable $k$-clause $3$SAT instance has,
modulo a fixed natural number $q$, a nonzero
number of non-solutions. Such counting checks are not known to be in the
polynomial hierarchy, thus the membership 
test for fixed $\omega$-languages
is represented by a complexity class possibly larger than
\textbf{NP} or \textbf{PH}. $\Box$
\end{remark}

\noindent
One-in-three-SAT is the following modification of $3$SAT problem.
Given a set of clauses, where each clause contains at most three
literals, is there a truth assignment to the variables such that
in each clause exactly one literal is true?
Note that the version of one-in-three-SAT where all the literals used
are positive literals and each variable occurs at most
three times in the formula is also NP-complete.
Note that, assuming ETH, this version of
the one-in-three-SAT problem requires time
$2^{\Omega'(m)}$, where $m$ is the number of variables.

\begin{theorem} \label{th:ufaomega}
(a) Assuming the Exponential Time Hypothesis, checking whether an $n$-state
UFA defines an $\omega$-word in a fixed $\omega$-regular
language $\cal L$
takes at least time $2^{\Omega'((n \log n)^{1/3})}$.

(b) Checking whether an $n$-state
UFA defines an $\omega$-word in a fixed $\omega$-regular
language $\cal L$
can be done in $2^{O((n \log^2 n)^{1/3})}$ time.
\end{theorem}
\begin{proof}
(a) Consider an instance of one-in-three-SAT formula (where all the literals
are positive, and each variable occurs at most three times)
is given, which contains $m$ variables and $m'$ clauses,
$c_1,c_2,\ldots,c_{m'}$.

The instance will be coded into a UFA such that,
$\omega$-word generated by the UFA will have
$1111 \cdot \{0100,0010,0001\}^* \cdot 1111$ as a substring
iff the given one-in-three-SAT instance has a solution.

Consider the string $\tau$ of length $4(m'+1)$ over the alphabet consisting of $0,1$ and the set of variables.
The first four bits of $\tau$ are $1$ and then there
are $m'$ blocks of four characters each. The $j$-th block
is $0xyz$ (respectively $00xy$), if the $j$-th clause $c_j$ contains three variables 
$x,y,z$ (respectively contains two variables $x,y$).
The overall $\omega$-word
generated by the UFA will be an infinite concatenation of $\tau$, where
the variables in $\tau$ are replaced by the truth values assigned to them
via some truth-assignment to the variables as described below.

The construction is similar to various previous constructions.
Let $h= \lceil m/\log m \rceil$, where the rounded down value
of $\log m$ is used. Consider distinct primes $p_0, p_1, \ldots, p_h$ which
are at least $m$ and at most $\Theta(m)$. Note that there exist
such primes by the prime number theorem.
Assign $\log m$ variables to each of the primes $p_1, p_2, \ldots, p_h$
(the set of variables assigned to different primes are pairwise disjoint).
The prime $p_0$ is used for a different purpose. 
For the prime $p_i$, make a cycle $C_i$ of length $4(m'+1) \cdot p_i$.
For $\ell < 4(m'+1)$, the cycle $C_i$ has $4(m'+1) \cdot k+\ell$-th state as accepting iff
(i) the $\ell$-th character of string $\tau$ 
is $1$ and $i=0$ or (ii) the $\ell$-th character of $\tau$
is a variable $x$ assigned to $p_i$ and
$x$ is assigned truth value true by the $k$-th truth
assignment (modulo $2^{\log m}$) to the variables assigned to $p_i$.
Note that $p_i \geq 2^{\log m}$, so all possible truth assignments
to the variables assigned to $p_i$ are covered.
The $\omega$-word generated by the above UFA has a substring
$1111 \cdot \{0100,0010,0001\}^* \cdot 1111$
iff the one-in-three-SAT formula instance is satisfiable by having exactly
one true literal in each clause.
Note that the automaton described above is a UFA as, if
$\ell$-th character of $\tau$ is $1$, then only cycle $C_0$
can accept strings of length $\ell$ (modulo $4(m'+1)$).
If the $\ell$-th character of $\tau$ is $x$, then only the cycle
$C_i$, where $p_i$ was assigned variable $x$ can accept
the strings of length $\ell$ (modulo $4(m'+1)$).

Note that each cycle is of length $\Theta(m^2)$. Thus,
the size $n$ of the above UFA is $\Theta(m^3/\log m)$.
Thus, $m=\Theta(n \log n)^{1/3})$.
Assuming ETH, The above construction thus shows
that $2^{\Omega'((n \log n)^{1/3})}$ is a lower bound on the computation
time for checking whether a UFA generates an $\omega$-word which is
a member of the fixed $\omega$-regular language
$$    
   (\{0,1\}^4)^* \cdot 1111 \cdot \{0100,0010,0001\}^+ \cdot
    1111 \cdot \{0,1\}^\omega
$$    
(b) The upper bound on time required for the problem is $2^{O((n \log^2 n)^{1/3})}$.
This can be done by
converting the UFA into a DFA and simulating the Muller automaton
of the given $\omega$-regular language on the DFA until one knows which
states of the Muller automaton occur infinitely often.
Details are omitted as this is similar to
the proof of Theorem~\ref{th:nfaomega} part (b).
\end{proof}

\section{Summary and Conclusion}

\noindent
This paper studies the complexity of operations on finite automata
and the complexity of their decision problems when the alphabet is unary
and $n$ is the number of states of the finite automata considered.
The following main results are obtained:

(1) Equality and inclusion of NFAs can be decided within time
$2^{O((n \log n)^{1/3})}$. The previous upper bound on time
was $2^{O((n \log n)^{1/2})}$ as given by Chrobak (1986), using DFA conversion,
and this bound was not significantly improved since then.

(2) The state complexity of operations of UFAs (unambiguous finite automata)
increases for complementation and union at most by quasipolynomial.
However, for concatenation of two $n$-state UFAs, the worst case is a UFA
of at least $2^{\Omega((n\log^2 n)^{1/3})}$ states. Previously the upper bounds
for complementation and union were exponential and the 
exponential type lower bound
for concatenation was not known.

Decision problems of finite formulas on $n$-state UFAs using
regular operations and comparision,
in the worst case, require an exponential-type of time assuming
the Exponential Time Hypothesis. This complexity goes down to
quasipolynomial time in the case that the concatenation of languages
is not used in the formula.
Merely comparing two languages given by $n$-state
UFAs in the Chrobak Normal Form is in \textbf{LOGSPACE}. For general UFAs,
one can do comparing, union, intersection, complement and Kleene
star in \textbf{POLYLOGSPACE}.

(3) Starting from this research, it is shown that there are
certain $\omega$-regular languages of infinite words such that deciding whether
an NFA / UFA generates an infinite word in that given 
$\omega$-regular language
is almost as difficult as constructing the DFA of that language from
the given automaton. Here the infinite binary word generated by a
finite automaton has as bit $m$ a $1$ iff the finite automaton
accepts a word of length $m$ -- as the alphabet is unary, one could
also write ``accepts the word of length $m$''.

The main results of this paper were presented
at the conference FSTTCS 2023~\cite{CDGHJSST23}.

\ifx\arxivy\noy

\newpage

\section*{Research Highlights}

\noindent
Highlights should inform about achievements of the paper in three to five
statements each not being longer than 85 letters including blanks.

\begin{enumerate}
\item Comparing unary NFAs in time exponential in the third root of $n \log n$;
\item Quasipolynomial bound on size of complement and union of unary UFAs;
\item Exponential in third root of $n$ lower bound
      for unary $n$-state UFA concatenation;
\item ETH-based lower bounds for many computational problems of unary automata.
\end{enumerate}
\fi

\end{document}